# Title: An algebraic thixotropic elasto-visco-plastic constitutive equation describing pre-yielding solid and post-yielding liquid behaviours

Lalit Kumar
*Department of Energy Science and Engineering, Indian Institute of Technology Bombay, Mumbai 400076, Maharashtra, India*

**Abstract:**

Formulating an appropriate elasto-visco-plastic constitutive equation is challenging, especially for a model accurately describing pre-yielding solid and post-yielding liquid behaviours. A few models tried to explain both behaviours simultaneously. Oldroyd's 1946 formulation was one of the first models explaining it, however, assumptions of a simple linear elastic and quasi-static deformation before yielding made his model idealistic. At the same time, the quasi-static pre-yielding deformation assumption open-up the possibility for the consideration of pre-yielding viscous and plastic deformation when quasi-static conditions are not fulfilled. Most of the earlier models followed Oldroyd's pre-yielding linear elastic assumption. Here, we discuss the structural parameters based thixotropic non-linear elasto-visco-plastic constitutive model valid for reversible (finite thixotropic time scale) and irreversible (infinite thixotropic time scale) thixotropic materials. In this work, we have considered non-linear elastic and plastic behaviours before yielding. Despite being a simple algebraic equation, our model appropriately explains both the viscosity plateau at low shear rates and the diverging zero shear rate viscosity, using the same parameters but different shear histories. Our model also predicts experimentally observable transient shear banding due to micro-structure breakage by shear rejuvenation and steady-state shear banding due to aging. Furthermore, our model predicts initial gel structure (waiting time) dependent stress overshoot during shear rate startup flow, different stress hysteresis in shear-rate ramps, sudden stepdown shear rate test results, and viscosity bifurcation during creeping flow phenomena effectively. Depending on shear histories, our model at the steady state reduces to either Bingham, Herschel Bulkley type or Newtonian fluids model. Our model requires only four parameters for the irreversible and five parameters for the reversible thixotropic-elasto-visco-plastic (TEVP) model obtainable from the rheometer test, compared to parameters required ranging from six to thirteen by the existing TEVP rheological model. Our model favourably predicts a series of recent experimental results. The current framework has the potential to provide a possible physical interpretation of the Bingham model.It also has the capability to predict a delayed flow start for an appropriate structure degradation kinetic.

1. **Introduction**

Soft matter such as complex fluids consisting of colloidal assemblies(1–3), microgel(4–7), emulsion(8,9), foams(10,11), and non-Brownian suspension(12,13) often has micro-structure with varying length and strength spanning over complete volume. These materials are generally classified as structured fluids and often show thixotropic elasto-visco-plastic (TEVP) (finite thixotropic time scale) or irreversible TEVP (infinite thixotropic time scale, also referred to as rheomalaxis fluids(14)) characteristics. Many industrial and natural materials come under these classifications, such as crude oil, paints, battery slurries, toothpaste, concrete, ink, mineral suspensions, adhesives, foodstuffs, personal care products, blood, and mining, coal and metal slurries. These materials show micro-structure weakly bonded together, often by weak Van der Waals forces, to successfully resist a small applied shear but de-structure under higher shear. These materials can or can't reform the micro-structure after removing stress or lowering the shear rate, depending on TEVP or irreversible TEVP characteristics, respectively. The strength of the micro-structure depends on the aggregation mechanisms and interaction forces between their constituents. Micro-structure assembly's varying length and strength often result in non-linear elastic and viscous behaviour.

Oldroyd(15), in his seminal work, extended the Bingham yield stress model(16) by considering elastic deformation before yielding. Oldroyd considered quasi-static conditions before yielding, neglecting any pre-yielding frictional/viscous loss. However, in most practical cases, the material undergoes a finite shear rate for a considerable time, even when the applied stress is lower than the yield stress(17). When the applied stress is lower than the yield stress, the shear rate reduces continuously to a negligible value after an extended period (18,19). These findings have been discussed and debated in detail(17,18,20–24). One set of arguments led by Barne and coworkers (18,20) claims the non-existence of yield stress and refers to it as a measurement artefact. They concluded this based on the existence of a finite shear rate in the case when the applied stress is less than the yield stress. Other sets of researchers argue that the measurement of yield stress requires the patience of the investigators(17,21,23). Moller and coworkers(17) show that if applied stress is less than the yield stress, the shear rate can eventually reach zero after a significant time delay. Hence, it can be inferred that, indeed, the yield stress is a reality for some materials. Previous to Moller and coworkers(17), yield stress has been argued as engineering reality(21), empirical reality(22), or even sociological reality(23), which requires the patience of investigators. Barnes and Walters(18) show a viscosity plateau at very low shear rates to justify their argument of no yield stress. However, Moller and coworkers(17) showed an increase in viscosity as a function of the long waiting time. An ideal TEVP model should explain both the observation of Barnes and Walters and Moller et al. at suitable time scales. We will demonstrate that our model predicts viscosity plateau at low shear rates in transient conditions. However, using the same parameters, our model predicts diverging steady-state viscosity at zero shear rate.

In literature, the thixotropic elasto-visco-plastic constitutive relation is sometimes approximately by considering time-dependent yield stress, thus ignoring initial elastic behaviour(25–29). To improve the yield stress based thixotropic model, researchers considered a structure dependent pre-yielding linear elastic behaviour, an extension of the Oldroyd(15) model (7,30–34). Very few TEVP models start with the viscoelastic model and plasticity and thixotropy are added into these models (35–38). Depending on the starting base model, de Souza Mendes and Thompson(39) classified the thixotropic elasto-visco-plastic constitutive equation into two types. The type I model starts with a viscoplastic model, often the Bingham or Oldyrod model, to which elastic and thixotropic effects are added(7,25,29–34,40–42). Whereas the type II model begins with the viscoelastic model, often the Maxwell or Jeffery model, to which plasticity and thixotropy are added(35–38). Recently, Larson and Wei(43) has comprehensively reviewed the thixotropic models. Hence, we briefly summarise the important work relevant to our modelling. Readers are advised to see Larson and Wei's review paper for a detailed review of the thixotropic model.

Mujumdar et al.(30) listed earlier the thixotropic model and improved the existing model by considering a continuous change from pre-yielding elastic behaviour to post-yielding plastic behaviour. They successfully predicted a non-monotonic stress-strain curve in shear startup flow. However, their stress-strain curve remains non-differentiable. Furthermore, in their case, viscosity increases as the structure degrades, which is contrary to experimental observations. Dullaert and Mewis(33) improved the TEVP model by considering viscosity due to suspension/structure fluid interaction as well. They were able to predict experimental results for changes in shear rate for buildup and breakage cases. Armstrong et al.(34) further argued that Dullaert and Mewis's model is unable to explain large amplitude oscillatory shear (LAOS) flow, shear flow reversal, etc. They showed that the Mujumdar et al.(30) model, modified by Armstrong et al.(44), appropriately predict the abovementioned phenomena. However, this model uses a large number of parameters, i.e., thirteen. de Souza Mendes and Thompson(35,37,39,44) developed a thixotropic elasto-visco-plastic rheology model based on the Jeffery (Oldyord-B) model. They tried to give a physically meaningful rheological model, a type II model. They further used their model for predicting transient flow behaviour as well. They argued that Bingham rheology lacks well physical meaning and was formulated to capture steady-state flow behaviour(39). However, their model is based on the Jeffery model, and it has characteristics of the Maxwell model. Due to this, it is unable to predict true yield stress for a finite value of parameters. The same can be predicted using their model with infinite suspension viscosity, in such cases, their model reduces to the Kelvin-Voigt model. However, the use of infinity as a parameter value limits the scope of the rheological model. Recently, Wei et al.(31) proposed a rheological model by introducing a spectrum of structure parameters for ideal thixotropy (only viscosity). They showed that the introduction of multiple structure parameters could successfully capture the non-monotonic evolution of stress or shear rate. Furthermore, they incorporated viscoelasticity together with multiple structure parameters(45) and combined it with the isotropic kinematic hardening (IKH) model(46). They explained that their 12 parameters model shows important rheological features like viscoelasticity and stretched-exponential thixotropic relaxation in step rate or step stress tests, a non-monotonic thixotropic response in intermittent flows, kinematic hardening, and linear viscoelasticity before yielding. Furthermore, Wei et al. used their model to study shear rate inhomogeneities and shear banding phenomena(42). Recently, Wang and Larson(7) formulated a simple thixotropic constitutive model with elastoplastic stress and a smoothly decreasing modulus near a solid boundary for predicting shear banding, a step change in the shear rate, etc. They emphasized the importance of a simple model for predicting complex rheological phenomena.

Thus, it becomes important to consider modelling TEVP constitutive equation using a different approach instead of adding different concepts together, leading to a large number of parameters and a complex constitutive model. Here, we consider total strain administering the structure breakdown while formulating a simple model capable of predicting a range of complex rheological behaviour for the different fluids. However, net changes in the micro-structure depend on the applied shear. Total strain along with shear rate has a true history of shearing, and earlier atomic theory and microstructure-based models were also formulated using total strain(47,48). Zaccone et al.(47) developed a simple non-linear stress-strain constitutive model based on the atomic theory of elasticity of amorphous solids. Furthermore, Laurati et al.(48)

utilized the confocal microscopy imaging technique to quantify mean square displacement and the number of particles that remain in the nearest neighbour for a long time. They predicted the rheological behaviour from the microscopic structure, supplemented with static structural information. Their model consists of elastic and viscous stresses as a function of total strain, which reduces to Newtonian flow like the Kumar et al.(38) model for a large value of strain. The present model considers the possibility of local yielding before global yielding, where the potential energy is low. According to Coussot and Rogers(49), the distributions in sizes, shapes, charge, and other physical characteristics result in localised deformation first in regions of shallower local potential energy than elsewhere. Hence, it can be inferred that the plastic deformation in the microstructure is a continuous process even in the elastic dominant regions, and in some places, plastic deformation may start, while in other areas elastic deformation continues. The breakage of micro-structure can be considered to span over a large deformation due to variations in the local energy potentials. Kumar and co-workers(38,50) theoretically argued the possibility of the continuous destruction of micro-structure upon deformation. While formulating a rheological model for irreversible TEVP fluids accounted for the possibility of viscous and plastic deformation before yielding, as most of the time deformation field doesn't obey the quasi-static condition as prescribed by Oldyord(15). The start of plastic deformation does not cease the elastic deformation, but the material yield surface continues to evolve in such a way that yield stress keeps on increasing until it reaches maximum, similar to isotropic hardening materials. After reaching the maxima, the breakage of the elastic mico-structure results in the reduction of yield stress. Subsequently, Dimitriou and Mckinley(41,46), while formulating the IKH based model for predicting the rheology of waxy crude oil, utilised total deformation inferring from the solid plasticity theory. Furthermore, Coussot and Rogers(49) provided experimental evidence showing both elastic and plastic deformation before yielding. Hence, they argued in favour of continuous structure breakage instead of a sudden collapse of the microstructure. Moreover, many materials are known to show non-zero storage and loss modulus during the small amplitude oscillatory shear (SAOS) test(51–53), indicating the possibility of energy dissipation before yielding. Hence, using these pieces of information, we build an elasto-visco-plastic-based constitutive relation consisting of elastic, viscous, and plastic deformation before yielding, which is valid for both reversible and irreversible thixotropic fluids.

We tried to provide a physical meaning of our model, and at the same time, our model predicts solid as well as liquid behaviours for a finite value of parameters. Despite being a simple algebraic equation, our model requires fewer parameters than the existing one requiring parameters ranging from six to thirteen (7,7,30,31,34,37,45,46,54). Our model considers, while a part of the micro-structure breaks and losses elastic deformation (local yielding), at the same time, other parts of the micro-structure continue to deform elastically and vice versa. In our model, elastic, plastic, and viscous deformation co-occurs. Hence, the overall deformation is responsible for both elastic and plastic deformation. In our model, initially, elastic deformation starts with zero-plastic strength. As deformation increases, elastic strength decreases, and plastic strength increases. In some cases, all elastic strength converts into plastic strength, and the model reduces to the Bingham equation. Our model can capture phenomena like irreversibility in the micro-structure changes, transient and steady-state shear banding, and prediction of shear stress drop in response to a sudden shear rate drop which is also used to distinguish thixotropy from viscoelastic (55,56). The prediction of true yield stress by the Jeffery and Oldyord-B model (39), due to the inherent characteristic of the Maxwell-based model, becomes difficult. At the same time, viscoplastic models like Bingham(16) or Oldyord(15) as base models result in stress discontinuity at the yield point(57). Our model can handle these shortcomings and does not have stress discontinuity at the yield point. In addition to these phenomena, our model successfully explained stress-hysteresis during shear ramps, waiting time-dependent stress overshoot during shear rate startup flow, and transient and steady-state shear banding phenomena. It also has the potential to explain phenomena like non-monotonic viscosity bifurcation during creeping flow, delayed yielding, etc. To explain the phenomena mentioned above, we have examined different structural kinetics equations. We have also analysed the effect of different structure evolution kinetics on the rheological behaviour of our model. Furthermore, we have shown that our model qualitatively predicts the recent experimental results of Dimitriou and Mckinley(22), Zhao et al.(50), Dinkgreve et al.(58), Serial et al.(59), Mendes et al.(60), Datta et al.(61), etc. We also qualitatively compared and analysed our results with the existing modelling results. The present model resembles the features of de Souza Mendes's(35) model during the transient flow of virgin gel and the characteristic of Dimitriou and Mckinley's(22) model during the transient flow of already broken gel (slurry state of the gel). Finally, we have shown that depending on shear history, our model at the steady state reduces to either Bingham, Herschel Bulkley type or Newtonian fluids model.

## 2. Gel degradation Kinetics

A generalized shear history-dependent thixotropic elasto-visco-plastic constitutive relationship can be written as

$$\tau = \mu_s \dot{\gamma} + \tau^{te} = \mu_s \dot{\gamma} + \mu_g(\lambda, \dot{\gamma}, \gamma, t) \dot{\gamma} + G(\lambda, \dot{\gamma}, \gamma, t)(\gamma_e + \gamma_p)$$
$$= \mu_s \dot{\gamma} + \mu_g(\lambda, \dot{\gamma}, \gamma, t) \dot{\gamma} + G(\lambda, \dot{\gamma}, \gamma, t)\gamma \qquad (1)$$

Where $\dot{\gamma}, \gamma, \gamma_e, \gamma_p, t$, and $\tau$ are the rate of deformation, total deformation, elastic deformation, plastic deformation, time, and deviatoric stress, respectively. Furthermore, $\mu_s$ is solvent viscosity, $\mu_g$ is shear history-dependent gel viscosity, and G is shear history-dependent elastoplastic modulus having information on elastic and plastic behaviours of the material. $\tau^{te}$ is generally modelled as Maxwell visco-elastic model (e.g., Jeffrey model). However, in the present work, we assumed the Kelvin-Voigt-based model (Fig. 1) for $\tau^{te}$ with structure-dependent elastic and viscous modulus. The schematic diagram shown in Fig. 1 consists of a Kelvin-Voigt-type mechanical circuit in parallel with a dashpot representing viscous dissipation in the liquid/solvent phase. In our schematic diagram, the Kelvin-Voigt diagram develops an extra frictional body in series with a spring body as the structure degrades. In our case, the elastic strength (elastic modulus) decreases as deformation increases, but elastic stress keeps increasing. The elastic stress increases until the stress reaches a maximum (i.e., a static yield stress, which can also refer to as global yielding). The decrease in the elastic modulus is due to the breakage of the local elastic bonds, allowing local plastic deformation. The loss in elastic strength is a gain in plastic strength. This indicates that much before global yielding, local yielding occurs in the regions with the lowest potential energy. Subsequently, the material yield surface continues to evolve in such a way that some parts of the material continue to deform elastically and others plastically. Here, the yield surface initially evolves in such a way that the yield stress and elastic deformation both increase until global yielding. However, the elastic deformation at some local points may continue without an increase in total elastic stress. In our model, elastic strength decreases, and plastic strength increases as deformation increases. In some cases, all elastic strength seems to convert into plastic strength, and the model reduces to the Bingham equation (Fig. 1). Further quantification of elastic and plastic deformation, before and after reaching the global yield point (static yield stress), is required, especially to explain the cases of a step change in stress. Maxwell model is usually used to explain viscoelastic liquids, and the Kelvin-Voigt model is used for viscoelastic solids. However, we will see later how adding structure-dependent Kelvin-Voigt coefficients together with frictional body enables our model to explain initial solid and subsequent liquid behaviour. The structure parameter $\lambda$ dependent on $\dot{\gamma}, \gamma$, and $t$.

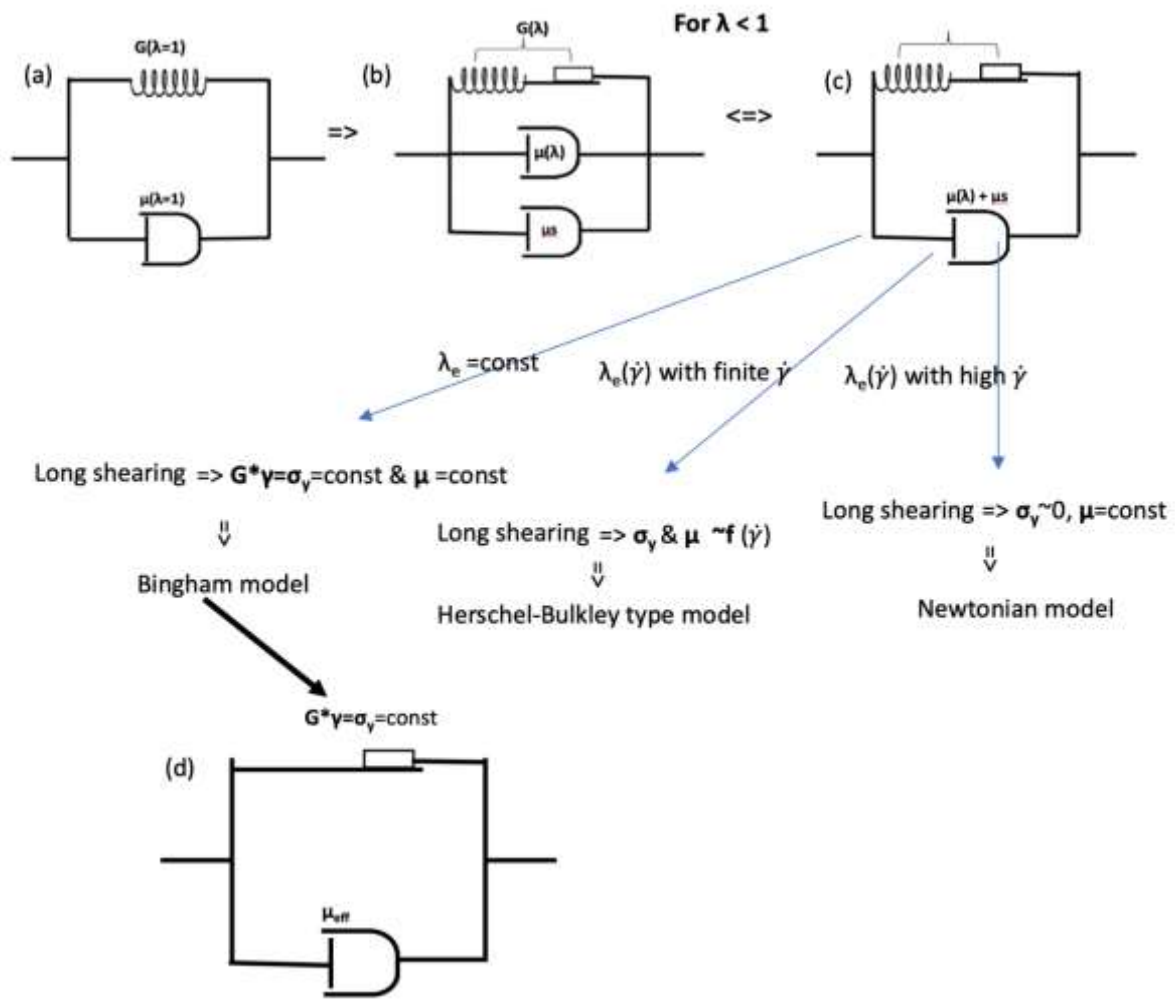

Fig. 1. Schematic diagram of our model where all the deformation is assumed to participate in the elastic-plastic and viscous phenomena, (a) shows that initially, our system behaves like Kelvin-Voigt viscoelastic material, (b) shows as deformation induced in the system plasticity develop in the system and both liquid and microstructure undergo the same deformation and total deformation contributes to the elastic force by modifying thixotropic modulus, (c) simple representation of (b) where both the viscosity is combined as an equivalent viscosity, and (d) represent a special case of elasto-visco-plastic where after certain deformation thixotropic modulus varies such that the product of thixotropic modulus and deformation becomes constant (i.e., Bingham model).

### 2.1 First-order gel degradation Kinetics

First order gel degradation model consisting of aging and shear-rejuvenation by following Moore(62) can be written as

$$\frac{d\lambda}{dt} = \frac{1-\lambda}{T_0} - m\dot{\gamma}\lambda \qquad (2)$$

Where $T_0$ is the thixotropic time scale, and $m$ is the gel degradation rate constant. In the above equation, aging is assumed to be proportional to $(1-\lambda)$, i.e., the structure required to be formed before reaching zero shear rate minimum potential energy state at a given temperature. Where $\lambda = 1$ is assumed to be the normalised structure parameter value for fully structured gel (i.e., the gel is aged for a long time and able to minimize potential energy at that temperature). From Eq. (2), the steady-state dynamic equilibrium structure parameter for a given shear rate can be obtained as

$$\lambda_e = \frac{\frac{1}{T_0}}{\frac{1}{T_0}+m\dot{\gamma}} = \frac{1}{1+mT_0\dot{\gamma}} \qquad (3)$$

Thus, the structure parameter $\lambda$ varies from 1 to $\lambda_e$, where $\lambda_e$ is a result of the dynamic equilibrium between aging and shear rejuvenation. Even in the absence of aging, the breakage of the micro-structure will depend on the shearing strength and $\lambda_e$ could have a finite value. This is related to the strength of micro-structure vs stress developed by shearing. The choice of constant $\lambda_e$ is possible using Kee et al.(63) model. In the case of a very large value of the thixotropic time scale or in the case of a very high shearing rate $\lambda_e$ approaches zero.

A general solution to Eq. (2) can be written as

$$\lambda = \lambda_e + (\lambda_0 - \lambda_e)e^{-(m\dot{\gamma}t)}e^{-t/T_0} \qquad (4)$$

Although the above equation is derived from a simple constant shear rate flow, we can generalise it by first substituting $\dot{\gamma}t = \gamma$ (which is strictly valid only for the case of constant shear rate) and then claiming the absolute value of strain is complex flow-dependent deformation. And it is also valid in more complex flows such as constant or variable shear rate/stress driven flow. We will further discuss this for other gel degradation kinetic models, where this generalisation is more obvious. The strain, $\gamma$, can be directly related to true creep compliance, which requires a solution of mass and momentum conservation equations together with the respective constitutive equation. For specific fluid and flow fields, generalised creep compliance can be obtained using the corresponding mass and force balance equation, as explained by Kumar et al.(38,64). Hence, the structure parameter for both shear rate and shear stress-driven flow can be written as

$$\lambda = \lambda_e + (\lambda_0 - \lambda_e)e^{-m\gamma}e^{-t/T_0} \qquad (5)$$

A similar relationship can be written for variable shear rates using other forms of evolution of a structure parameter(63,65). The first-order structure evolution equation proposed by Kee et al.(63) can be written as

$$\frac{d\lambda}{dt} = -b\dot{\gamma}(\lambda - \lambda e) \tag{6}$$

This results in

$$\lambda = \lambda_e + (\lambda_0 - \lambda_e)e^{-(m\dot{\gamma}t)} \tag{7}$$

Furthermore, we can solve Eq. (6) by first considering either constant or variable shear rate and then substituting $\dot{\gamma} = \frac{d\gamma}{dt}$. This will allow us to solve the structure parameter $\lambda$ as a function of deformation.

$$\lambda = \lambda_e + (\lambda_0 - \lambda_e)e^{-m\gamma} \tag{8}$$

By comparing Eqs. (7) and (8), it appears that we have substituted $\dot{\gamma}t = \gamma$. However, as discussed earlier that the formulation of Eq. (8) allows us to consider the structure model in a more complex fluid flow scenario, i.e., in addition to constant shear rate flow Eq. (8) can be used in other complex flow scenarios such as flow driven by applied stress or pressure. Here, equation (8) is valid for any type of flow, either having a constant or variable shear rate. Since $\lambda_e$ is a parameter, it enables different functions of $\lambda_e$ depending on the aim of the model. Coussot et al. (65) give the structure evolution equation in the presence of aging and shear rejuvenation as follows.

$$\frac{d\lambda}{dt} = \frac{1}{T_0} - m\dot{\gamma}\lambda \tag{9}$$

The above structure kinetic equation assumes the same shear rejuvenation term as in Eq. (2), but in this case, the rate of aging is assumed to be constant and inversely proportional to the thixotropic time scale. If we analyse the above equation in the absence of shear rejuvenation, it reveals that the structure-building process is linear with time, and it does not dependent on the present state of microstructure or concentration of crystallising components present in the solution. This is a very unlikely practical condition considering different mechanisms evolve at different stages of structure building. Hence, this model appears simplistic, however, despite simplified assumption, many important rheological learning(29,65–67) have been achieved using Eq. (6). Furthermore, if we take $\lambda_e = 1/mT_0\dot{\gamma}$ in Kee et al.(63) models, then it becomes exactly same as the Coussot et al. (65) model.

Above equation with initial condition $\lambda = \lambda_0$, gives

$$\lambda = \lambda_e + (\lambda_0 - \lambda_e)e^{-(m\dot{\gamma}t)} \tag{10}$$

Equations (7) and (10) appear exactly the same, however, Eq. (10) is derived from the Coussot et al. model, and $\lambda_e = 1/mT_0\dot{\gamma}$ is well-defined and has a fixed value. On the other hand, in the Kee et al. model $\lambda_e$ act like a parameter, we will later demonstrate how different choices of $\lambda_e$ result in different familiar steady-state rheological models.

### 2.2 The $n^{th}$ and third-order gel degradation Kinetics

Now we will discuss the $n^{th}$ and third-order gel structure degradation model based on Kee et al.(63) models. The third-order model for waxy crude oil is proposed by Paso et al.(68) and further developed by Kumar and co-workers(38,50) as a function of strain while formulating rheological equations for irreversible TEVP materials. Paso et al. suggested that the structure parameter derived using the first-order kinetics decreases sharper than some industrial materials like waxy crude oil micro-structure. Hence, they proposed a third-order model with a long tail (i.e., gel degraded much slower when smaller micro-structures were left unbroken). In this model, the breakage is proportional to the difference between the current structure level and the equilibrium level to power three. Furthermore, we will examine later that the third order gel degradation kinetics with appropriate assumption provides the possibility of the popular yield stress model (e.g., Bingham, Hershel-Buckley-type models) at the steady state. Whereas the first-order model always reduces to Newtonian fluid at a steady state, as the first-order model predicts sharper gel degradation at a longer time. Similar to our first-order model, a constitutive equation as a function of total deformation was also formulated using a micro-model(47,48), which also predicts Newtonian behaviour at a steady state. For the $n^{th}$-order model, the evolution of structure parameters based on the Kee et al.(63) model is given by(68)

$$\frac{d\lambda}{dt} = -m\dot{\gamma}(\lambda - \lambda_e)^n, \quad \text{where } n \neq 1 \tag{11}$$

This results in

$$\lambda = \lambda_e + \frac{1}{((n-1)m\gamma + 1/(\lambda_0 - \lambda_e)^{n-1})^{\frac{1}{n-1}}} \tag{12}$$

For n=3 above equation becomes a structure evolution equation for the third-order model,

$$\lambda = \lambda_e + \frac{1}{(2m\gamma + 1/(\lambda_0 - \lambda_e)^2)^{\frac{1}{2}}} \tag{13}$$

For small values of $\lambda_e$ and $\lambda_0 = 1$, the above equation reduces to

$$\lambda \cong \lambda_e + \frac{1}{(2m\gamma + 1)^{\frac{1}{2}}} \tag{14}$$

The equilibrium structure parameter acts like a variable instead of a predefined parameter in Kee et al.'s (63) first-order model. We now want to combine the learning from the first order and try to incorporate additional decaying terms, which only depend on a thixotropic time scale. This will help in predicting the delayed start of elasto-visco-plastic fluids. This can be understood from Eq. (5), where initially, the structure degrades due to deformation by the applied forces. However, once elastic forces approximately balance the applied force, the deformation almost becomes stagnant. In the absence of further deformation, the structure degradation due to shear rejuvenation stops. However, once deformation becomes stagnant, the second term causing gel degradation depends on time and is influenced only by the thixotropic time scale. This may contribute towards further structure degradation, especially in the case where build-up will be negligible. This may eventually reduce the yield requirement and result in a delayed flow restart. Hence, similar to the first-order gel degradation model, we also multiply the time dependent degradation of the structure parameter by including the exponential decaying term as in Eq. (5). After multiplying the second term with exponentially decaying in time term, the structure parameter evolution equation becomes as follows.

$$\lambda = \lambda_e + \frac{e^{-t/T_0}}{(2m\gamma + 1/(\lambda_0 - \lambda_e)^2)^{\frac{1}{2}}} \qquad (15)$$

The above equation is referred to as third-order micro-structure degradation kinetics with the possibility of a delayed restart. And the algebraic equation (15) with constant $\lambda_e$ might be useful for predicting delayed yielding. In such a case, there will not be any build-up in the absence of a shear rate, similar to the irreversible TEVP case, but the second term will reduce with time.

Table1. List of different gel degradation models with their corresponding dynamic equilibrium structure expression.

| Model derived or approximated from | Expression | $\lambda_e$ |
|---|---|---|
| Kee et al. (57) | $\lambda = \lambda_e + (\lambda_0 - \lambda_e)e^{-(m\gamma)}$ | Parameter |
| Coussot(65) | $\lambda = \lambda_e + (\lambda_0 - \lambda_e)e^{-(m\gamma)}$ | $\dfrac{1}{mT_0\dot{\gamma}}$ |
| Moore(62) | $\lambda = \lambda_e + (\lambda_0 - \lambda_e)e^{-m\gamma}e^{-t/T_0}$ | $\dfrac{1}{1 + mT_0\dot{\gamma}}$ |
| Kumar et al.(38) | $\lambda = \lambda_e + \dfrac{1}{(2m\gamma + 1/(\lambda_0 - \lambda_e)^2)^{\frac{1}{2}}}$ | Parameter chosen as 0 |
| Proposed in this work to predict delayed yielding | $\lambda = \lambda_e + \dfrac{e^{-t/T_0}}{(2m\gamma + 1/(\lambda_0 - \lambda_e)^2)^{\frac{1}{2}}}$ | Parameter |

Before developing a rheological model, we summarise gel degradation behaviour and its effect on the possible rheological model. We discuss here how the choice of gel degradation kinetics influences the characteristics of the rheological model. It will assist in choosing different gel degradation kinetic depending on the behaviour of the gel, especially to predict steady state results at low shear rate conditions. Table 1 shows various expressions of structure parameters discussed in this report. Most previous works argue that shear rate or shear stress are responsible for gel breakage. However, in the case of applied stress conditions also gel responds with the development of shear rate, including in the true yielding materials with applied stress lower than the yield stress. Hence, the model consisting of either shear rate dependent breakage or shear stress breakage has to overcome yield strain. Therefore, we converted our gel structure parameter into strain dependent model, which has information on stress and shear rate histories. The conversion from shear rate-dependent gel degradation to strain-dependent structure parameter becomes evident in the case of the Kee et al. model. Hence, we argue that in Kee ta al.'s model, the dynamic equilibrium structure parameters can be selected to replicate other models mentioned in table 1. Further, it can be noticed that shear rate dependence in the structure parameter comes from dynamic equilibrium structure parameters (and subsequently in the rheological model). Most previous work uses first-order gel degradation kinetics given by either Kee et al., Coussot et al., or Moore or their extension.

We used third-order gel degradation kinetics as it allows us to formulate a rheological constitutive equation, which predicts yielding with or without aging at steady state conditions (detailed discussion will be in subsequent sections). In our constitutive model framework, the first-order gel degradation kinetics will always predict Newtonian flow at a steady state, as structure degradation is fast. In all cases, the structure parameter has two terms; the first is called the dynamic equilibrium structure parameter, which for thixotropic fluids, depends on balancing between aging and shear rejuvenation. For ir `reversible thixotropic fluids, a constant $\lambda_e$ can be taken. For irreversible thixotropic fluids, $\lambda_e$ depends on the highest shear rate experienced by the fluids as it will determine the final microstructure of the broken gel. The final structure is the result of the balancing force across the micro-structure by applied shearing to attractive internal forces. And for a particular shearing field, the micro-structure will be fixed, leading to a constant $\lambda_e$. This is particularly valid when the structure is unable to reform once broken by shearing. In our case, the second term of the structure parameter gives shear rate independent breakage. In this work, we have used two different values of $\lambda_e$, i.e., constant and $1/(1 + mT_0\dot{\gamma})$, for irreversible and reversible thixotropic fluids, respectively. However, our model currently does not support changing rheological behaviour from non-aging to aging. This is possible by supplementing our model with a micro-structure model like population balance(69). The shearing strength capable of converting non-Brownian micro-structure into Brownian constituent enables the conversion of a non-aging model into an aging model.

Finally, we have included a time-dependent term in the third-order model similar to Moore's (62) first-order model, the last row in table 1 with a constant $\lambda_e$. This helps explain delayed yielding when the applied stress is less than static yield stress. F When the applied stress is less than yield stress for true yielding material, the initial deformation remains less than the corresponding yield strain. In such cases, the shear rate dependent $\lambda_e$ shows an aging effect, and the material regains initial strength. However, for a constant value of $\lambda_e$, the material will remain deformed (with some elastic and some local plastic deformation) without yielding globally, but the broken structure will not reform. The initial deformation will decrease the second term of the structure parameter, however, the combined value of the structure parameter may be large enough for elastic force to balance the applied force. Hence, we have included a time-dependent term similar to Moore's (62) model to explain delayed yielding. The inclusion of $e^{-t/T_0}$ terms can further decrease the structure parameter value with time, resulting in a reduction in the yield stress requirement. This can be referred to as delayed yielding, as these types of materials have a very large thixotropic time scale. Hence, a constant value of $\lambda_e$ will be important in predicting initial non-aging behaviour and shear rate dependent $\lambda_e$ for aging. Recently, Dinkgreve et al.(58) have shown that initial non-aging Carbopol solution becomes aging after a long duration of stirring. The nature of the broken micro-structure is found to be responsible for the conversion of non-Brownian gel (non-aging) into Brownian gel (aging). To deal with aging and non-aging material behaviour simultaneously, our model requires the inclusion of a micro-structure model like the population balance model(69).

### 3. Structural parameter-based Thixotropic Elasto-viscoplastic model

The proposed thixotropic elasto–visco-plastic stress model is given by

$$\tau = (\mu_s + \mu_g * \lambda)\dot{\gamma} + G_0 * \lambda^3 * \gamma \tag{16}$$

In the above model, the liquid part of the gel or solvent viscosity ($\mu_s$) is assumed to be constant, whereas both the gel viscosity ($\mu_g * \lambda$) and the gel elasto-viscoplastic modulus ($G_0 * \lambda^3$) depend on the state of micro-structure and micro-structure interaction with liquid. The structure parameter has two separate contributions: the dynamic equilibrium constant, which depends on a thixotropic time scale, the degradation rate constant, and the shear rate. The second term depends on total deformation and is independent of the shear rate. It can be seen from Eq. (3) that in the case of low shear rates for a material having a low thixotropic time scale, the gel structure remains intact as $\lambda_e$ approaches one. Furthermore, the elastic and plastic effects depend upon the three-dimensional microstructure, whereas viscosity can be associated with the orientation of the structure, and one dimension is enough to influence the viscosity. Hence, the gel viscosity has first-order dependence on the structure parameter in our model, whereas the elastic modulus has third-order dependence. It is also consistent with the experimental observation where elastic modulus decreases faster than viscosity for most materials. Kumar and co-workers(38,50) model used the difference between the current structure level and dynamic equilibrium structure to characterise the viscosity and elasto-plastic modulus, whereas in the present model both the elasto-plastic modulus and viscosity depend on the present state of the structure level. The assumption of viscosity and elastic-plastic modulus dependence on the present structure level makes dynamic equilibrium viscosity and elastic-plastic modulus of the gel non-vanishing in some cases, resulting in a true and aging dependent yield stress model at steady state conditions. However, the final broken gel structure is expected to depend on the shearing strength. It has also been reported that structure breakage below the Kolmogorov length scale is difficult to achieve, even in the highly sheared turbulent regime. Hence, it is safe to assume that some crystals may survive even in the case of high shearing. Finally, it can also be understood that the broken structure may or may not be able to form connecting structure in the dynamical conditions, depending on the strength and period of the applied shearing. The final broken structure can be classified into three types; (I) a connected broken non-aging sub-micro-structure able to resist some finite applied stress, (II) a connected/non-connected broken micro-structure shows an aging effect via Brownian motion, and (III) a broken structure neither has connecting nor Brownian sub-micro-structure. In the steady state flow condition, the first type of material shows simple yielding behaviour, the second type shows aging-dependent yield stress and viscosity at a low shear rate, and in the last case, the material shows Newtonian flow behaviour. Hence, the final broken slurry's micro-structure determines the slurry state's rheology as either the Bingham, the Herschel-Buckley-like, or Newtonian behaviour. To explain the step-down stress behaviour, the determination of recoverable and unrecoverable deformation becomes essential. This can be done following Rogers and co-workers (49,70,71). However, in this work, we will not go into the details of recoverable and unrecoverable deformation, which is critical mostly in explaining a step-down in shear-stress results. This work focuses on the rheological behaviour under constant or variable shear rates, including cyclic shear rates test and constant or increasing shear stress cases.

We used the $\lambda$ value from Eq. (13) to calculate the thixotropic elastic-plastic modulus of Eq. (16). Furthermore, in the case of a very small value of $\lambda e$ only the linear term of $\lambda e$ is considered, and the rest of the terms are neglected in our model. This will result in the following TEVP model.

$$G = G_0 \left( \lambda_e^3 + 3\lambda_e \frac{1}{(2m\gamma + 1/(\lambda_0 - \lambda_e)^2)^{\frac{1}{2}}} \left( \lambda_e + \frac{1}{(2m\gamma + 1/(\lambda_0 - \lambda_e)^2)^{\frac{1}{2}}} \right) + \frac{1}{(2m\gamma + 1/(\lambda_0 - \lambda_e)^2)^{\frac{3}{2}}} \right)$$

$$\sim G_0 \left( 3\lambda_e \frac{1}{(2m\gamma + 1/(\lambda_0 - \lambda_e)^2)} + \frac{1}{(2m\gamma + 1/(\lambda_0 - \lambda_e)^2)^{\frac{3}{2}}} \right) \quad (17)$$

The above equation makes the stress function as follows

$$\tau = \mu_s \dot{\gamma} + \mu_g \left( \lambda_e + \frac{\lambda_0}{(2m\gamma\lambda_0^2 + 1/(1-\lambda_e/\lambda_0)^2)^{1/2}} \right) \dot{\gamma} + G_0 \left( \frac{3*\lambda_e*\lambda_0^2}{(2m\gamma\lambda_0^2 + 1/(1-\lambda_e/\lambda_0)^2)} + \frac{\lambda_0^3}{(2m\gamma\lambda_0^2 + 1/(1-\lambda_e/\lambda_0)^2)^{3/2}} \right) \gamma \quad (18a)$$

When $\lambda_0 = 1$ and $\lambda_e$ small above equation can be simplified as,

$$\tau = \mu_s\dot{\gamma} + \mu_g\left(\lambda_e + \frac{1}{(2m\gamma+1)^{1/2}}\right)\dot{\gamma} + G_0\left(\frac{3*\lambda_e}{(2m\gamma+1)} + \frac{1}{(2m\gamma+1)^{3/2}}\right)\gamma \tag{18b}$$

Similarly, a constitutive equation for first-order gel degradation kinetics is written as follows

$$\tau = \mu_s\dot{\gamma} + \mu_g(\lambda_e + (\lambda_0 - \lambda_e)e^{-m\gamma})\dot{\gamma} + G_0\,(3\lambda_e\lambda_0^2 e^{-2m\gamma} + \lambda_0^3 e^{-3m\gamma})\,\gamma \tag{19a}$$

When initially microstructure is completely formed ($\lambda_0 = 1$) i.e., it reaches to the static equilibrium at that temperature, the above equation reduces to

$$\tau = \mu_s\dot{\gamma} + \mu_g(\lambda_e + (1 - \lambda_e)e^{-m\gamma})\dot{\gamma} + G_0\,(3\lambda_e e^{-2m\gamma} + e^{-3m\gamma})^3\gamma \tag{19b}$$

Similar constitutive equations as a function of total deformation were also formulated using a micro-model(47,48), which predicts Newtonian behaviour at a steady state similar to our first-order model.

### 4. Results and Discussion

#### 4.1 Steady-state flow behaviour, as a limiting case analysis

We first analysed three limiting cases, where the effect of the equilibrium structure parameter and shear histories on the slurry state viscosity and yield stress are evaluated. We studied how different expressions of $\lambda e$ control our rheological model at steady-state conditions.

**Limiting case (I)**, after a long time of shearing of non-aging material leads $\lambda e$=constant

$$\tau = \mu_g\lambda_e\dot{\gamma} + G_0\left(\frac{3*\lambda_e}{2m\gamma}\right)\gamma + \mu_s\dot{\gamma} = \mu'_{eff}\,\dot{\gamma} + \tau_y \tag{20}$$

This approximation converts the slurry state rheology of the thixotropy elasto-visco-plastic model into Bingham rheological model (Fig. 1). This may happen when a broken micro-structure has a non-aging connecting structure. Hence, due to the connecting micro-structure, it can resist some finite applied stress. However, the broken sub-micro-structure is large, so no Brownian motion effect is observed. The state of the broken micro-structure is determined by the balance between internal force within the micro-structure which keep the microstructure connected, and external shearing force. This will be a constant for non-aging material and related to the highest shearing experience by the materials.

**Limiting case (II)**, after a long time of shearing of aging material with any shear rate may lead to $\lambda e = c1 * \dot{\gamma}^{-c_2}$

$$\tau = \mu_g(c1 * \dot{\gamma}^{-c_2})\dot{\gamma} + G_0\left(\frac{3*c1*\dot{\gamma}^{-c_2}}{(2m\gamma+1)}\right)\gamma + \mu_\infty\dot{\gamma}$$

$$= \mu'_g\,\dot{\gamma}^{1-c_2} + \tau_y{'}\dot{\gamma}^{-c_2} + \mu_s\gamma = \mu_g(\dot{\gamma})\dot{\gamma} + \tau_y(\dot{\gamma}) \tag{21}$$

In this case, the final structure of the gel network and its orientation highly depend on the applied shear rate. This scenario arises when the broken micro-structure at a steady state shows aging behaviour. At a low shear rate, the connecting micro-structure reforms. In this case, our model reduces to a relation similar to Herschel Buckley's type of model with shear rate dependent yield stress.

**Limiting case (III)**, after a long time of shearing with a very high shear rate $\lambda e=0.0$ can be considered as no connecting structure may remain. In this case, the broken micro-structure also form a non-aging constituents, and hence

$$\tau = \mu_s \dot{\gamma} \tag{22}$$

In the case of very high shear rates, most of the network in the gel will be broken to such an extent that no flow directional-dependent orientation takes place. Furthermore, broken structures are unable to reform connecting structures by aging. Hence, the fluid may start behaving like a Newtonian fluid.

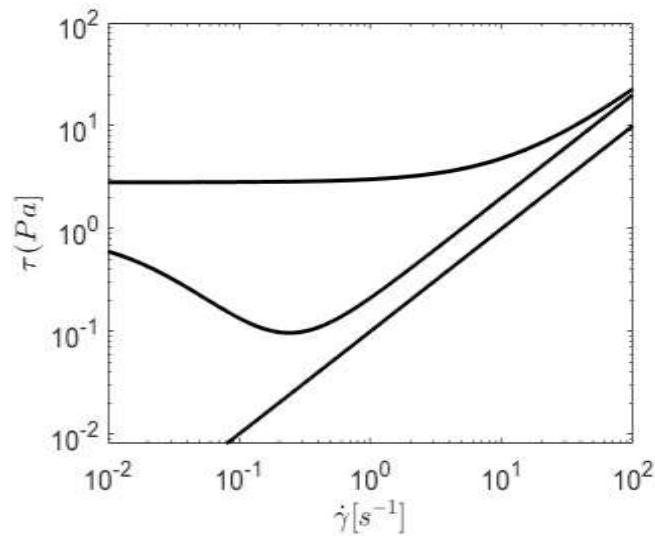

Fig. 2. Showing stress predicted by our model as a function of shear rate at steady state conditions for $\lambda e = constant = 0.0015$ as in the limiting case (I), $\lambda e(\dot{\gamma})$ with $T_0 = 2000$ s as in the limiting case (II), and $\lambda e=0$ as in the limiting case (III) other parameters used for cases corresponding to (I) and (II) are $\mu_s$=0.12 Pa s, $\mu_g$=0.5 Pa s, $G_0$=50000 Pa , m=40 and for case (III) , $\mu_s$=0.1 Pa s

Figure 2 shows the results of our model in the limiting steady-state cases as discussed above, i.e., $\lambda e = constant = 0.0015$ corresponding to the limiting case (I), $\lambda e(\dot{\gamma})$ corresponding to the limiting case (II) and $\lambda e=0$ corresponding to the limiting case (III). Recently, Van der Geest. et al.(72) performed experiments with six different crude oils and presented stress vs shear rate curves assumingly at steady-state. However, Van der Geest. et al. has two types of results, one is a verified steady state and another in an unverified steady state (which is indeed a transient state result). Our model predicts both types of Van der Geest. et al. results. The verified steady-state results show a very good match with our steady-state results, whereas the unverified steady-state results match our transient results. Their crude oil samples 3 and 5 show type (II)-behaviour, and other samples show type (I) behaviour. It is also consistent with the Mendes et al.(60) results. Mendes et al. showed that depending on the sample preparation method and composition of waxy crude, it can show thixotropic or irreversible thixotropic behaviour. For unverified steady-state results of Van der Geest. et al., we identify the flow condition after 30 or 5 minutes of shearing. To further illustrate this, if we consider the lowest shear rate in their work, which is $10^{-7}$ s$^{-1}$, for this shear rate material takes~100 hr to reach even yield strain, and hence it can't be referred to as steady-state results after 30 minutes. After 30 minutes, the sample develops 0.00018 units of strain only. Hence, this can be referred to as transient flow, and we will show later that it matches very well with our transient results.

The steady-state results can be helpful while selecting parameters for a transient model for different virgin gels. The choice of $\lambda e$ becomes clear for a material with known steady-state characteristics (aging vs non-aging). Hence, with prior information of the steady state flow behaviour, like for a simple yield stress (non-aging) flow, $\lambda e = constant$ can be used, similarly, in the case of aging dependent yielding at a low shear rate, the strain rate dependent $\lambda e$ can be used. For Newtonian steady-state flow behaviour, either $\lambda e = 0$ or an infinitely large value of the thixotropic time scale is required. A special case of our model when $\lambda e = 0$ is already formulated by Kumar et al.(38) and used for predicting flow restart(38,73) in a pipeline and transient slip phenomena(74). Subsequently, we will also discuss how to obtain other important parameters of our model from experimental results.

**4.2 Stress as a function of deformation in the case of a constant shear rate start-up flow**
This sub-section analyses the results of a constant shear rate start-up flow in a different condition and qualitatively compares our prediction with existing experimental results. Here, we will also demonstrate the utilization of quasi-static conditions, stress overshoot strain and stress values to obtain the gel degradation rate constant, *m,* and elasto-plastic modulus, *G$_0$*. We first compare our results with the experimental results of Zhao et al.(50) and Dinkgreve et al.(58), demonstrating stress overshoots during a constant shear rate startup flow. Fig. 3a shows the stress as a function of strain for waxy crude oil gel (an ITEVP fluid) during shear rate startup flow for various shear rates (50). For waxy crude oil known as ITEVP fluid, we have taken $T_0$->infinity, implying $\lambda_e=0$, as discussed earlier. For other parameters, we have utilized maximum stress information. The maximum stress value of Eq. (18b) for $\lambda_e=0$ at quasi-static condition is $G_0/(3^{3/2}m)$, where m= $1/\gamma_{max}$, $\gamma_{max}$ is the strain at which stress becomes maximum. Other parameters like $\mu_s$ and $\mu_g$ are chosen arbitrarily. The value of $\mu_s$ can be obtained experimentally by measuring steady-state viscosity at a very high rate. Whereas, $\mu_g$ can be obtained by measuring stress at low strain values using different shear rates, similar to the waiting time dependent results of Dimitriou and McKinley(46). However, obtaining these data requires carefully constructed experiments. The available results enable only *m* and *G₀* values, other parameters $\mu_s$ and $\mu_g$ are chosen arbitrarily. Equation (18) has three terms: the first term represents the liquid part viscosity, the second term represents the gel viscosity whose origin comes from liquid-microstructure interaction and microstructure-micro-structure dissipative interaction, and the third part is elastic-plastic component due to Van der Waals interaction/entanglement in the microstructure. Fig. 3b shows our model results which closely resemble the experimental results of Zhao et al.(50). It captures initial linear and nonlinear elastic behaviours, stress overshoot, and subsequent decrease in stress. Similarly, Fig. 3c shows the stress as a function of strain for stirred Carbopol sample, which shows the thixotropic effect, during shear rate startup flow for various shear rates (58). Dinkgreve et al.(58) suggested that a long duration of stirring breaks the Carpobol network to such an extent that it starts showing a thixotropic effect as the broken structure becomes Brownian. Eq. (18) at the maximum stress gives m= $1/\gamma_{max}$ for $\lambda_e=0$ and m~ $0.3975/\gamma_{max}$ for $\lambda_e=1$, when the maximum stress reaches quasi-statically. Dinkgreve et al.(58) used stirred Cabopol as a working fluid, which is also a weakly thixotropic fluid. For this result, we chose m=3 as the maximum stress between $\gamma = 0.2$ to $\gamma = 0.5$ (we choose $\gamma_{max} = 0.33$), this gives $G_0$=770 Pa in the case of irreversible thixotropic fluids. Here, we select $G_0$=600 Pa considering stirred Carbopol as weakly thixotropic and a fixed value of $\lambda_e=0.1$. We qualitatively predict all the characteristics, including stress overshoot (Fig. 3d). However, our model predicts smoother overshoot for all shear rates, compared to sharper overshoot found experimentally for some shear rates. Despite using nonlinear elastic-plastic modulus, we are able to predict initial apparent linear elastic behaviour (Fig. 3d). This happens as the gel degradation constant has a small value, and the value of elastic-plastic modulus remains the same for a small value of strain.

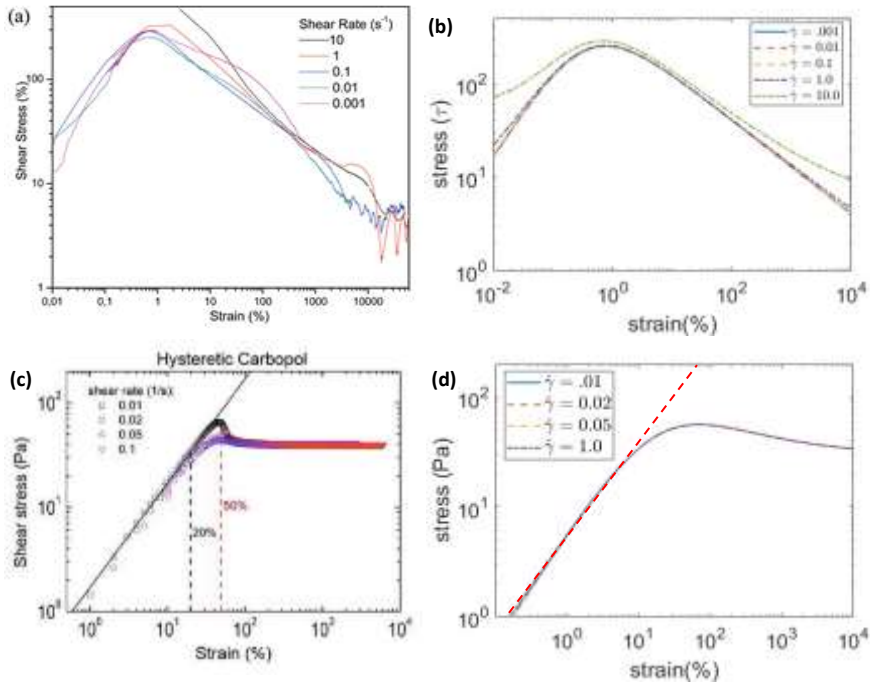

**Fig. 3.** Stress as a function of deformation for different values of shear rates using our rheological model given in Eq. (18) plotted for (a) experimental results of Zhao et al.(50), (b) our model prediction using $T_0 \to$ infinity 100000 s, m=100, $\mu_s$=0.05 Pa s, $\mu_g$=5 Pa s, $G_0$=50000 Pa, (c) experimental result of Dinkgreve et al.(58), and (d) our model prediction using $\lambda_e$=0.1, m=3, $\mu_s$=0.05 Pa s, $\mu_g$=0.05 Pa s, $G_0$=600 Pa.

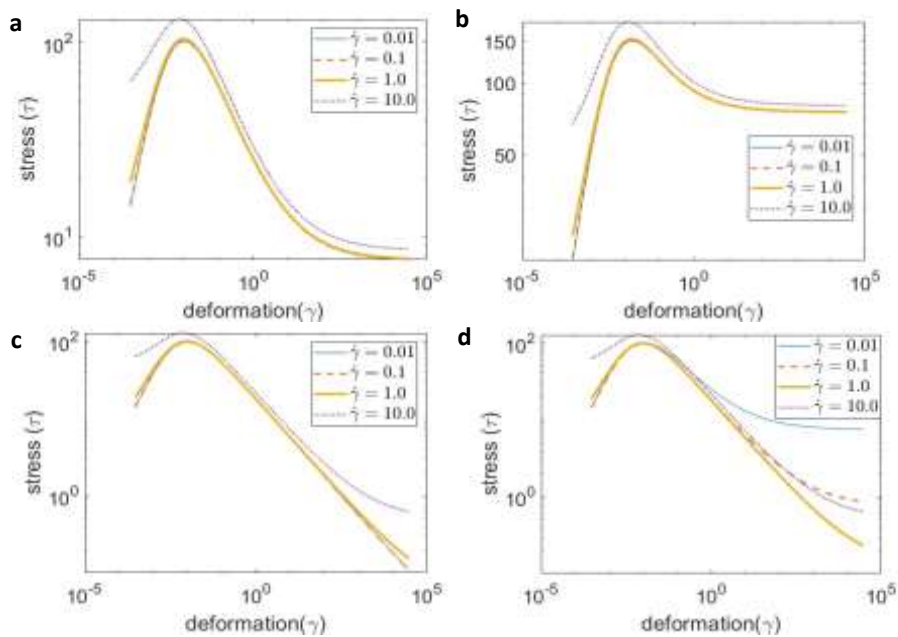

Fig. 4. Stress as a function of deformation for different values of shear rates using our rheological model given in Eq. (18) plotted for (a) $\lambda_e$=0.01, and (b) $\lambda_e$=0.1 using a constant value of $\lambda_e$ whereas for (c) $T_0$=100000 s, (d) $T_0$= 100 s for shear rate dependent $\lambda_e$, and other parameters m=100, $\mu_s$=0.05 Pa s, $\mu_g$=5 Pa s, $G_0$=50000 Pa are kept constant for all plots.

Figure 4 further demonstrates the capability of our model while predicting stress behaviour during different shear rate start-up flows. Figure 4 shows stress as a function of deformation at various shear rates for different values of dynamic equilibrium structure parameters ($\lambda_e$) while keeping other parameters in Eq. (18) same. Figures (4a) and (4b) are plotted for a constant value of $\lambda_e$, 0.01 and 0.1 respectively. In the case of constant $\lambda_e$, the gel viscosity remains independent of the shear rate, and hence, the viscous shear stress at the steady state increases linearly with an increase in the shear rate. For a constant $\lambda_e$, at a steady state, the elastic part of stress also remains independent of the shear rate. In the case of aging dependent yielding, $\lambda_e$ is taken as a function of the thixotropic time scale ($T_0$), structure breakage rate constant (m), and shear rate ($\dot{\gamma}$), as given in Eq. (12). Figure (4c) and (4d) is plotted for different thixotropic time scales, $T_0$=100000 s and $T_0$=100 s, respectively. In these cases, as $m$ is constant, $\lambda_e$ depends on $T_0$ and $\dot{\gamma}$. For a large value of the thixotropic time scale $T_0$, $\lambda_e$ quickly approaches zero as the shear rate increases, and the model reduces to an irreversible thixotropic model. It can be seen from Fig. 3c that at high shear rates, the total stress remains high. Total stress has a dominant contribution from viscous stress. However, Fig. 3d for $T_0$=100 s shows contrasting stress profiles. In this case, for lower shear rates, the build-up term dominates breakage, as the microstructure formation is relatively faster. The presence of microstructure results in a non-zero value of elastic stress. And the extent of the microstructure becomes more complex as the shear rate becomes low. This results in higher elastic stress for a lower shear rate, which dominates the total stress (Fig. 4d). However, interestingly, Fig. 4d shows higher total stress for the case of $\dot{\gamma}$=10 s$^{-1}$ compared to $\dot{\gamma}$=1 s. This is because viscous stress becomes significant in the case of a high shear rate. At a moderate shear rate, the microstructure is unable to build for elastic stress to dominate total stress, and at the same time, viscous stress also remains low. However, in case of a further decrease in the shear rate, the build-up of microstructure starts dominating breakage, resulting in a dominant elastic stress. Hence from figure 4d, it can be observed that total stress becomes the highest for the lowest shear rate ($\dot{\gamma}$=0.001 s), dominated by elastic stress. This trend changes, and stress becomes higher for $\dot{\gamma}$=10 s compared to $\dot{\gamma}$=1 s due to viscous stress contribution, which differs from low shear rate cases. Based on these results, we can define three possible yield stresses, as discussed by Chang et al. (75). According to them, deviation from linearity is the first type of yield stress referred to as elastic-limit yield stress. However, in our case, we assumed a non-linear elastic modulus. Despite elastic modulus being non-linear, the stress-strain curve initially shows linear-like behaviour. Hence, we can define the elastic-limit yield stress as the stress at which the zero strain slope starts deferring by more than 10%. The other two yield stresses definition are the maximum value of the stress in the stress-strain curve can be referred to as static yield stress, and vanishing shear rate stress as dynamic yield stress.

**4.3 Stress overshoot as a function of waiting time.**

Figure 4 shows the stress overshoot for a constant shear rate start-up flow. The stress overshoot has been reported for many materials, waxy crude oil(46,50,72,76), Carbopol microgel(77), polymer solutions(78,79), etc. For entangled polymeric materials(80,81), it has been argued that during initial deformation, the elastic modulus due to the entanglement of polymers increases until the yield point, resulting in an increase in stress. However, beyond the yield point, the disentanglement results in a decrease in stress. This entanglement and disentanglements around the yield point result in an overshoot of stress(80,81). The extent of stress overshoot for thixotropic fluids is observed to increase as a function of aging time (waiting time)(46). However, accurate prediction of this phenomenon using rheological modelling is challenging. We have predicted stress overshoot as a function of the initial state of micro-structure (Fig. 5) and compared our prediction with the experimental and IKH model results of Dimitriou and McKinley(46). Fig. 5a and 5b show the experimental and modelling results of Dimitriou and Mckinley. For our modelling, we choose a fixed value of $\lambda_e = 0.05$, and $m = 1/\gamma_{max}$ is taken, where $\gamma_{max}$ is the strain value at which stress becomes maximum. For calculating $G_0$, the maximum stress value and m are used. From Eq. 18a, it can be seen that $G_0$ is proportional to the initial structure, and the initial structure is a function of waiting time. Hence, as the initial structure build-up with time, the thixotropic modulus and maximum value of stress increase proportionally. The other parameter taken are $\mu_s=\mu_g=0.1$ Pa s. Our model results are given in Fig. 5c, which qualitatively predicts the experimental results. However, the maximum stress and corresponding strain appear higher than the experimental values. Here, while choosing $m$ and $G_0$, we selected the parameter values valid in the case of quasi-static conditions. Hence, we adjusted the value of $m$ and $G_0$. The experimental result in Fig. 5a shows a large variation in stress at no or minimal deformation (initial stage). Hence, it can be concluded that initial gel viscosity also plays a critical role while predicting stress at a low value of strain, as the applied strain rate is the same in all cases. It can also be observed from Eq. 18a that the gel viscosity is also proportion to the initial gel structure parameter constant. Hence, we also varied $\mu_g$ proportional to the increase in $G_0$, as it depends on aging time. The corresponding modelling results are shown in Fig. 5d, which is much closer to the experiment results compared to Dimitriou and McKinley's IKH model results (Fig. 5b). From Fig. 5a and Fig. 5d, it appears that aging dependent model may predict better results at a larger time (i.e., a larger value of strain). For a larger strain value, the sample that initially aged for a lower period of time may show more aging effect, and aging dependent yield stress values may converge with each other. For better prediction of waiting time dependent stress overshoots, more details information on shear histories during the breakage stage and rheological behaviour without waiting time is required.

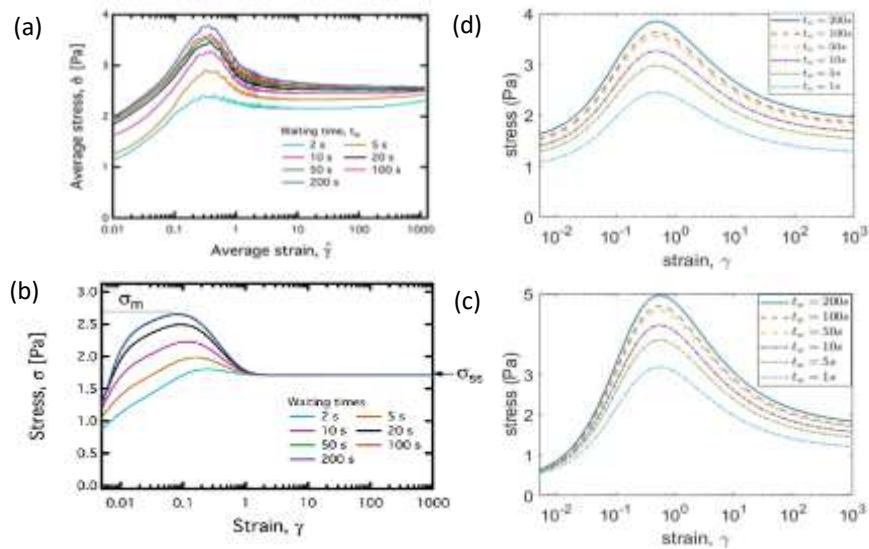

Fig. 5. Stress as a function of strain for different values of the aging time (a) Dimitriou and McKinley's experimental results(46), (b) Dimitriou and McKinley(46) IKH model prediction, (c) our prediction for $m = \frac{1}{\gamma_{max}} = 2.5$ and $\tau_{max} = \frac{G_0}{3^{\frac{3}{2}}m} => G_0 = 3^{\frac{3}{2}} m \, \tau_{max}$ and (d) condition is in (c) strictly valid for quasi-static condition so in this case $m = 3.1$, $\mu_g$=0.3*$G_0/G_{0,min}$ Pa s ( where, $G_{0,min}$ is the smallest elastic modulus value corresponding to minimum waiting time) and $G_0 = G_{0(c)} * \left(\frac{2.5}{3.1}\right)^2$ are considered, other parameters $\lambda_e = 0.05$, $\mu_s$=0.1 Pa s, $\dot{\gamma} = 2 \, s^{-1}$ are kept constant for all cases.

## 4.4 Stress as a function of the strain rate:

In section 2, we have discussed how to account for a varying shear rate on the microstructure breakage. In this sub-section, we will discuss the influence of a varying shear rate on the build-up terms. This becomes important, especially when the shear rate decreases with time and, subsequently equilibrium parameter ($\lambda_e$) increases. For a fixed value of strain rate, $\lambda_e$ remains a constant as in Eq. (10). However, it is important to note that often in the practical case, the rate of strain varies with time, and accordingly, $\lambda_e$ also changes especially for aging materials. In a practical case, it is challenging to use a thixotropic model design for a fixed shear rate, where the shear rate varies with time. In this section, we first explain how to use our model for varying shear rate cases. Equation (10) is a solution of the gel degradation kinetic Eq. (9). Thus, every time we solve Eq. (9), the outcome depends on the initial condition ($\lambda_0$) and parameters like ($T_0$, m and $\dot{\gamma}$) via $\lambda_e$. For a particular gel at constant temperature m, $T_0$ is assumed to be constant. Hence, in such cases, $\lambda$ depends on the time and $\dot{\gamma}$. One has to solve Eq. (9) each time for a varying shear rate with changing initial conditions, as previous shear history changes the initial micro-structure state. The current state of the micro-structure is used as an initial condition for the next stage (i.e., at other shear rates). Thus one will get Eq. (10) for each value of the shear rate with different initial conditions. Here, we assume that the gel has degraded for $dt_1$ time with a shear rate $\dot{\gamma}_1$ and then $\dot{\gamma}_2$ shear rate is applied for $dt_2$ time. This results in new initial conditions for $\dot{\gamma}_2$ shear rate, as given by

$$\lambda_{0new} = \lambda_e(\dot{\gamma}_1) + (\lambda_0 - \lambda_e(\dot{\gamma}_1))e^{-(m\dot{\gamma}_1 dt_1)} \tag{23}$$

So a new value of the structure parameter is given by
$$\lambda = \lambda_e(\dot{\gamma}_2) + (\lambda_{0new} - \lambda_e(\dot{\gamma}_2))e^{-(m\dot{\gamma}_2 dt_2)}$$
$$= \lambda_e(\dot{\gamma}_2)(1-e^{-(m\dot{\gamma}_2 dt_2)}) + \lambda_e(\dot{\gamma}_1) e^{-(m\dot{\gamma}_2 dt_2)} (1-e^{-(m\dot{\gamma}_1 dt_1)}) + \lambda_0 e^{-(m(\dot{\gamma}_1 dt_1 + \dot{\gamma}_2 dt_2))} \tag{24}$$

Similarly, for the next time step, we can calculate the new value of the structure parameter as
$$\lambda = \lambda_e(\dot{\gamma}_3)(1 - e^{-(m\dot{\gamma}_3 dt_3)}) + \lambda_e(\dot{\gamma}_2)(1-e^{-(m\dot{\gamma}_2 dt_2)}) e^{-(m\dot{\gamma}_3 dt_3)} + \lambda_e(\dot{\gamma}_1) (1-e^{-(m\dot{\gamma}_1 dt_1)}) e^{-(m(\dot{\gamma}_1 dt_1 + \dot{\gamma}_2 dt_2))} + \lambda_0 e^{-(m(\dot{\gamma}_1 dt_1 + \dot{\gamma}_2 dt_2 + m\dot{\gamma}_3 dt_3))} \tag{25}$$

And if we continue further until the nth shear rate, we will get structure parameters as follows
$$\lambda = \lambda_e(\dot{\gamma}_n)(1 - e^{-(m\dot{\gamma}_n dt_n)}) + \lambda_e(\dot{\gamma}_{n-1})(1-e^{-(m\dot{\gamma}_{n-1} dt_{n-1})}) e^{-(m\dot{\gamma}_n dt_n)} + \lambda_e(\dot{\gamma}_{n-2}) (1-e^{-(m\dot{\gamma}_{n-2} dt_{n-2})}) e^{-(m(\dot{\gamma}_n dt_n + \dot{\gamma}_{n-1} dt_{n-1}))} + \ldots\ldots + \lambda_0 e^{-(m(\dot{\gamma}_1 dt_1 + \dot{\gamma}_2 dt_2 + m\dot{\gamma}_3 dt_3 + \cdots))} \tag{26}$$

Where '*n*' denotes how many different shear rates have been applied. Equation (26) can be written in the form as follows.

$$\lambda = \lambda_{equ} + \lambda_0 e^{-(m\gamma)} \tag{27}$$

Where for the first initial shear rate ($\dot{\gamma}_1$) applied for $dt_1$ time (i.e., for n=1 indicating single shear rate)

$$\lambda_{eqn}(1) = \lambda_e(\dot{\gamma}_1) * \left(1 - e^{-(m\dot{\gamma}_1 dt_1)}\right) \tag{28}$$

For the second shear rate ($\dot{\gamma}_2$) applied for $dt_2$ time (i.e., for n=2)

$$\lambda_{eqn}(2) = \lambda_e(\dot{\gamma}_1) * \left(1 - e^{-(m\dot{\gamma}_1 dt_1)}\right) e^{-(m\dot{\gamma}_2 dt_2)} + \lambda_e(\dot{\gamma}_2)\left(1 - e^{-(m\dot{\gamma}_2 dt_2)}\right) \tag{29}$$

For the nth shear rate (n>2) for time $dt_n$

$$\lambda_{equ}(n) = \lambda_{eqn}(n-1) * e^{-(m\dot{\gamma}_n dt_n)} + \lambda_e(\dot{\gamma}_n)\left(1 - e^{-(m\dot{\gamma}_n dt_n)}\right) \tag{30}$$

Equivalently for 3$^{rd}$ order gel degradation kinetics structure parameter value can be written as

$$\lambda = \lambda_{eqv} + \frac{\lambda_0}{(2m\gamma + 1)^{\frac{1}{2}}} \tag{31}$$

Where for the first initial shear rate ($\dot{\gamma}_1$) applied for $dt_1$ time

$$\lambda_{eqn}(1) = \lambda_e(\dot{\gamma}_1) * \left(1 - \frac{\lambda_0}{(2m\dot{\gamma}_1 dt_1 + 1)^{\frac{1}{2}}}\right) \tag{32}$$

For the second shear rate ($\dot{\gamma}_2$) applied for $dt_2$ time

$$\lambda_{eqn}(2) = \lambda_e(\dot{\gamma}_1) * \left(1 - \frac{\lambda_0}{(2m\dot{\gamma}_1 dt_1 + 1)^{\frac{1}{2}}}\right) \frac{\lambda_0}{(2m\dot{\gamma}_2 dt_2 + 1)^{\frac{1}{2}}}$$
$$+ \lambda_e(\dot{\gamma}_2)\left(1 - \frac{\lambda_0}{(2m\dot{\gamma}_2 dt_2 + 1)^{\frac{1}{2}}}\right) \tag{33}$$

For nth subsequent shear rate applied for time $dt_n$

$$\lambda_{eqn}(n) = \lambda_{eqn}(n-1) * \frac{\lambda_0}{(2m\dot{\gamma}_n dt_n + 1)^{\frac{1}{2}}}$$
$$+ \lambda_e(\dot{\gamma}_n)\left(1 - \frac{\lambda_0}{(2m\dot{\gamma}_n dt_n + 1)^{\frac{1}{2}}}\right) \tag{34}$$

Now we have used Eq. (31) in Eq. (18) and plotted stress as a function of the shear rate as shown in Fig. 6. For this figure, we have changed the shear rate linearly such that the maximum shear rate reaches in $2*10^6$ s for Fig. 6a, and in $2*10^4$ s Fig. 6b.

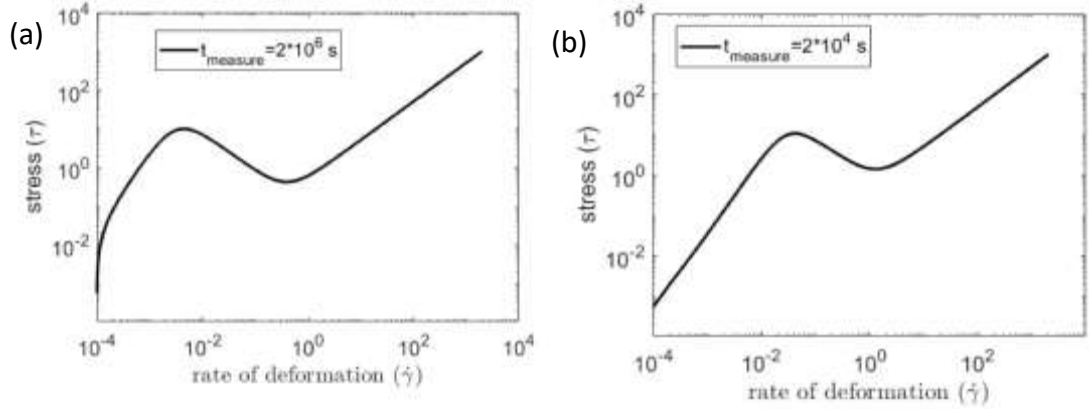

Fig. 6. Stress as a function of the rate of deformation for parameters $T_0$= 1000 s, $\mu_s$=0.5 Pa s, $\mu_g$=10 Pa s, m=100, $G_0$=5000 Pa, where the rate of strain increased linearly in such a way that the maximum strain rate reaches in $2*10^6$ s for (a), and in $2*10^4$ s for (b).

Literature has reported a complex stress behaviour (transient shear banding) during a shear rate sweep test for thixotropic elasto-visco-plastic fluids(46,61,72,76,82,83). Most of the existing thixotropic models are unable to predict the correct stress behaviour during the shear rate sweep test. It requires a prediction of the initial solid-like elastic stress jump, creep regime, followed by a smooth increase in stress until maxima dominated by elastic effect, subsequent decreases in the stress as microstructure degrades, and finally increase in the stress due to viscous-dominated effect at high shear rates. Dimitriou and McKinley(46), using KIKH (Kelvin isotropic-kinematic hardening) model, is able to qualitatively predict a decrease in stress from maxima, followed by an increase in stress as a function of shear rates. Wang and Larson(7) also reported a similar prediction using a boundary-induced modulus gradient. Both of these models were formulated to capture steady-state rheology of slurry, not virgin gel, and hence unable to predict initial elastic jump as discussed by Chang et al.(75). In contrast, de Souza Mendes's (35) thixotropic model at steady-state was able to predict an initial increase in stress as shear rate increased. We will discuss the comparison in more detail later. We have used Eq. (31) in Eq. (18) and plotted stress as a function of the strain rate, the strain rate is changed linearly to reach the highest value of strain rate in $2*10^6$ s for Fig. 6a, and in $2*10^4$ s for Fig. 6b. In case of a simple yield stress fluid, initially the stress increases with strain rate. Then after certain strain rate, it remains constant for a small period of time (i.e., referred to as yielding) before increases linearly as the strain rate increases significantly. For the thixotropic elasto-visco-plastic fluids, it has been reported that initially, stress increases as the shear rate increases similar to a simple yield stress fluid without a thixotropic effect. However, once it reaches the yield point, instead of constant stress, it starts decreasing until a certain strain rate before increasing again(75). Our model is able to predict the initial increase in the stress as the shear rate starts increasing, creeping regime, maxima, followed by a decrease in shear stress for the intermediate value of the shear rate, and finally, an increase in the stress for a high value of shear rate. We found that stress behaviour is a strong function of measurement time, especially in the initial stages. For Fig. 6 shear rate is increased linearly ($\dot{\gamma} = kt$), and hence deformation as a function of time is given by $\gamma = kt^2/2$. By the time the deformation reaches a value (say $\gamma = 5 * 10^{-3}$) where elastic component significantly contributes to the total stress, the strain rates become $10^{-3} s^{-1}$ and $10^{-1} s^{-1}$ for Fig. 6a and 6b, respectively. Hence, Fig. 6a shows an initial elastic jump, however, Fig. 6b has viscous like behavior at low shear rates. For Fig. 6a, at a low shear rate, total stress has a significant contribution from the elastic part, however, once stress reaches a local maximum value, the elastic contribution starts decreasing. A decrease in stress is observed at intermediate values of strain rate due to the breakage of microstructure, by the time rate of strain reaches a significantly large value. This results in a significant decrease in the elastic strength of the material, and hence, the elastic stress decreases rapidly. At the same time, the shear rate does not increase significantly to increase or maintain the stress. Thus, the total stress starts decreasing, however, when the rate of strain increases further, viscous stress starts dominating total stress. In this regime, total stress starts increasing with the strain rate. In cases where the initial gel viscosity is high, the decrease in stress after reaching maximum is found to be low. Gel viscosity is a weaker function of the structure parameter. Thus for high gel viscosity cases, the viscous stress due to the increase in strain rate will compensate for the elastic stress losses. Prediction of stress behaviour for all values of shear rate further confirms the capability of our model in explaining different transient flow regimes, including the initial stage. In the case of continuous increase in the shear rates, a constant value of $\lambda_e$ as in Eqs. (13) and (27) predict a similar stress profile as in Fig. 4. This happens as both terms in Eqs. (13) and (27) decrease with an increase in the shear rate. In Eqs. (13) and (27), the first term represents the structure build-up which decreases as the shear rate increases, and the second term represents the structure breakage also decreasing with the time of shearing. However, for a cyclic shear rate or the step-down in the shear rate above analysis becomes essential. For step-down or decreasing shear rates, the first term in Eqs. (13) and (27) increase at low shear rates due to aging, while the second term decreases.

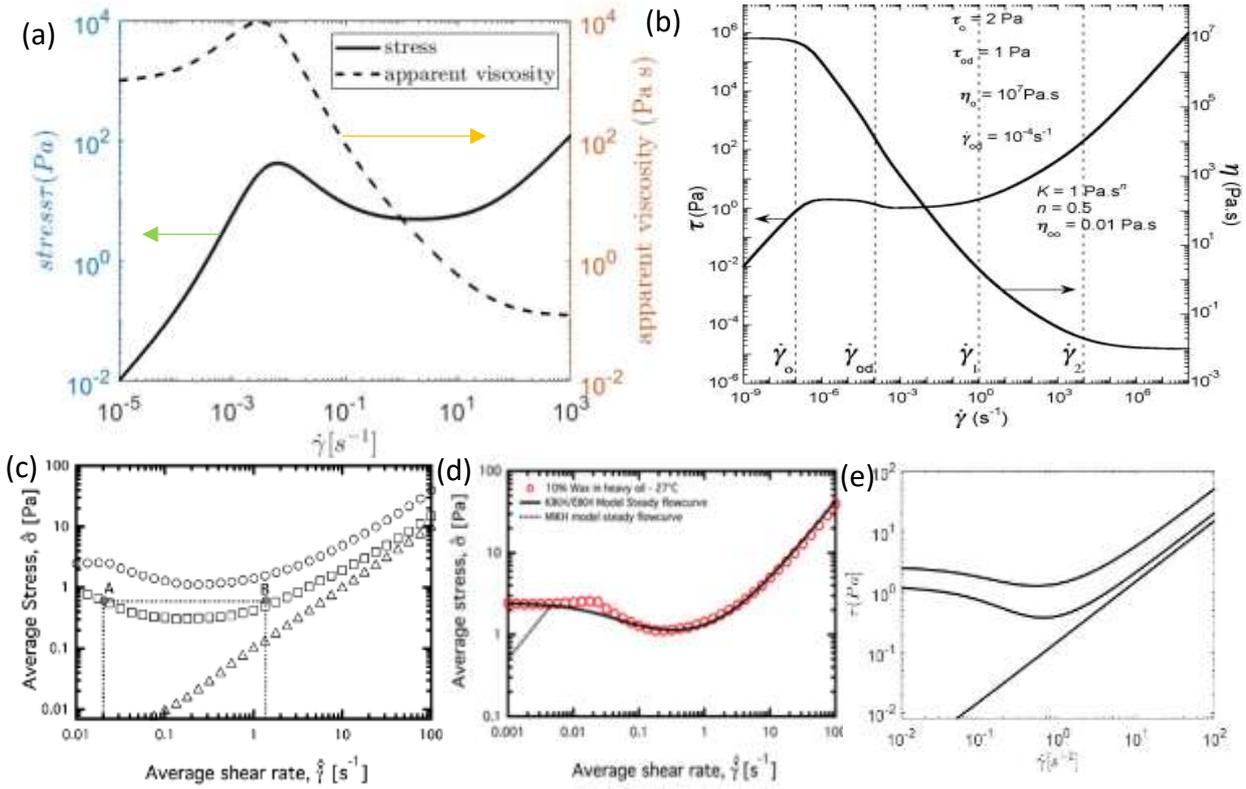

Fig. 7. Stress and apparent viscosity as a function of the strain rate (a) our modelling results for parameters $T_0$= 10000 s, $\mu_s$=0.1 Pa s, $\mu_g$=1000 Pa s, m=50, $G_0$=10000 Pa, where the rate of strain increased linearly in such a way that the maximum strain rate reaches in $2*10^5$ s, (b) de Souza Mendes(35) modelling results in the steady state condition, (c) Dimitriou and McKinley(46) experimental results, (d) Dimitriou and McKinley(46) modelling results for wax in oil, and (e) our modelling results predicting Dimitriou and McKinley(46) experimental results, using $T_0$= 20000000 s, $\mu_s$=0.5 Pa s, $G_0$=2000000 Pa for upper curve, $T_0$=5000000 s, $\mu_s$=0.2 Pa s, $G_0$=300000 Pa for middle curve and $\mu_s$=0.15 Pa s for lower curve, while $\mu_g$=1.0 Pa s, m=1 remains the same for all.

Furthermore, we have compared our results with the existing modelling results of de Souza Mendes(35) and the experimental and modelling results of Dimitriou and McKinley(46). de Souza Mendes model shows apparent yield stress behaviour (viscosity plateau followed by decrease in viscosity, Fig 7b) at a steady state. Whereas our model shows an apparent yield stress model in transient conditions (Fig. 7a) and yield stress behaviour with or without aging, Newtonian behaviours at steady state conditions (Fig. 7e), similar to Dimitriou and McKinley's experimental and modelling results (Fig. 7c and 7d). Dimitriou and McKinley's experimental result shows aging-dependent yielding behaviour at low shear rates, and Newtonian behaviour for waxy crude oil and heavy mineral oil, respectively. Whereas Dimitriou and McKinley's modelling result (7d) is for a fluid that shows aging-dependent yielding behaviour at low shear rates.

First, we analyse similarities and differences between our and de Souza Mendes's models. In our case, the reason for transient behaviour comes from delayed elastic deformation for a small shear rate leading to a dominant initial viscous effect, in contrast, de Souza Mendes has a very high limiting viscosity at low shear rates. Due to a very small deformation, elastic deformation continues for a short period and is equivalently accounted as viscosity. After reaching a static yield value, both models show shear rejuvenation. In our model, both gel viscosity and elastic modulus decrease and in de Souza Mendes's model, the viscosity decreases. During the intermediate shear rate regime, our model shows a transition of flow from elastic dominated to viscous-dominated flow. At the same time, de Souza Mendes's model predicts constant stress for a subsequent interval of shear rates due to competition between shear rejuvenation and aging. In the transient virgin gel case, the structure does not break enough at this stage (for intermediate shear rates), resulting in little or no aging effect. In our case, we observed aging at a low shear rate while predicting a steady flow of broken gel (Fig. 7e). In the steady state case, aging happens as most of the gel has already undergone an extensive shear rejuvenation and the micro-structure break in the Brownian sub-structure. Hence, we also noticed aging behaviour at a low shear rate while predicting steady-state flow. At a higher shear rate, our transient and de Souza Mendes steady-state models show a viscous dominant regime and behave similarly. Hence, the nature of our transient stress curve for virgin gel appears similar to de Souza Mendes's steady state curve, but the dynamics are different for initial and intermediate shear rates. At steady state conditions, we were able to predict Dimitriou and McKinley's steady state results (Fig. 7c), including the effect of aging. In the case of a shear-dependent dynamic structure parameter with an infinitely large thixotropic time scale, the stress curve follows Newtonian flow characteristics at a steady state, similar to heavy mineral oil, as shown in Fig. 7c. However, in the presence of aging material reform microstructure resulting in zero-shear rate stress (aging dependent yielding characteristic, diverging viscosity). Hence, we have shown that our model predicts the transient behaviour of virgin gel like de Souza Mendes's steady-state model and the steady-state behaviour of broken gel similar to Dimitriou and McKinley aging dependent yielding.

**4.5 Stress-hysteresis during shear rate up and down test**

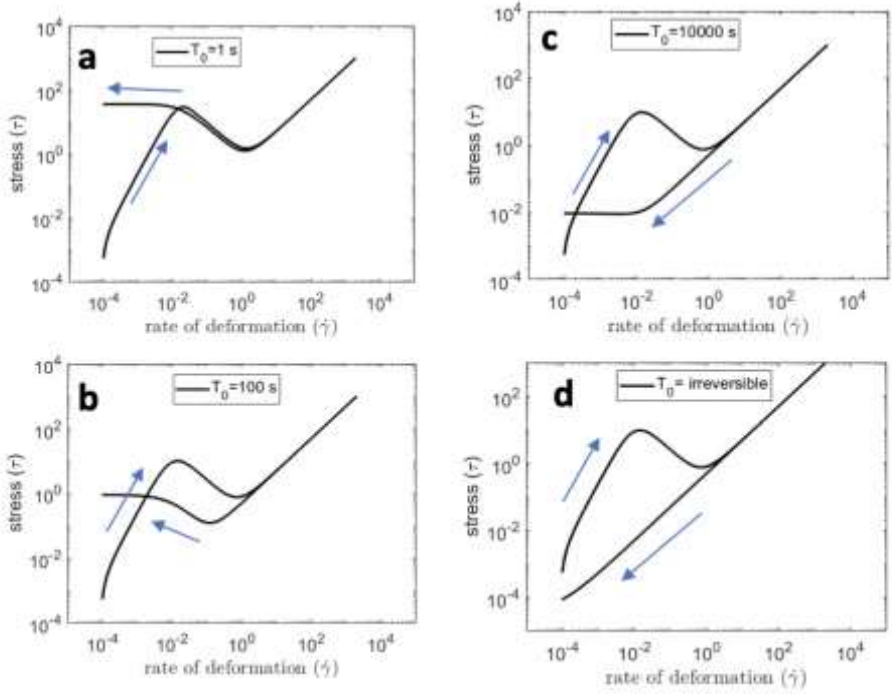

Fig. 8. Stress as a function of the rate of deformation for different values of thixotropic time scale (a) $T_0 = 1$ s, (b) $T_0 = 100$ s, (c) $T_0 = 10000$ s, and (d) $T_0 \to$ infinity, other parameters m=100, $\mu_s$=0.5 Pa s, $\mu_g$=5 Pa s, $G_0$=5000 Pa are kept constant for all plots.

Furthermore, many works of literature have reported stress-hysteresis for different materials during cyclic shear rate up and down sweeps test(7,19,46,75,84–90). Using our model, we have predicted stress-hysteresis during strain rate sweep test for irreversible (infinite thixotropic time scale) and reversible (finite thixotropic time scale) thixotropic materials. We have increased the shear rate linearly to a maximum value in 2000 s, and without waiting at the maximum value, again, the shear rate is decreased linearly to the initial point in 2000 s (Fig. 8). Fig. 8 presents different shear hysteresis as a function of the thixotropic time scale. Fig. 8d corresponds to a very large value of the thixotropic time scale, making the dynamic equilibrium structure parameter ($\lambda_e$) vanishing. It can also be referred to as hysteresis for irreversible thixotropic fluids. For irreversible thixotropic material, once the microstructure is broken, it cannot reform, and hence material cannot regain strength. Therefore, the stress in the up-sweep remains higher than the down-sweep stress, as similar hysteresis is reported by Mendes et al.(60) for irreversible thixotropic fluid like waxy crude oil. Furthermore, from Figs. 8a- 8c, it can be observed that as the thixotropic time scale becomes finite and increases further, the extent of the micro-structure recovery decreases at low shear rates. At a lower thixotropic time scale, the structure build-up is faster, leading to a rise in the viscosity and elasticity of the materials. Due to the increase in the viscosity and elasticity, the stress predicted for a lower value of the thixotropic time scale becomes more. For reversible thixotropic materials (aging materials), the stress at lower shear rates is experimentally observed higher during the shear down sweep compared to the shear up sweep(91), similar to our prediction.

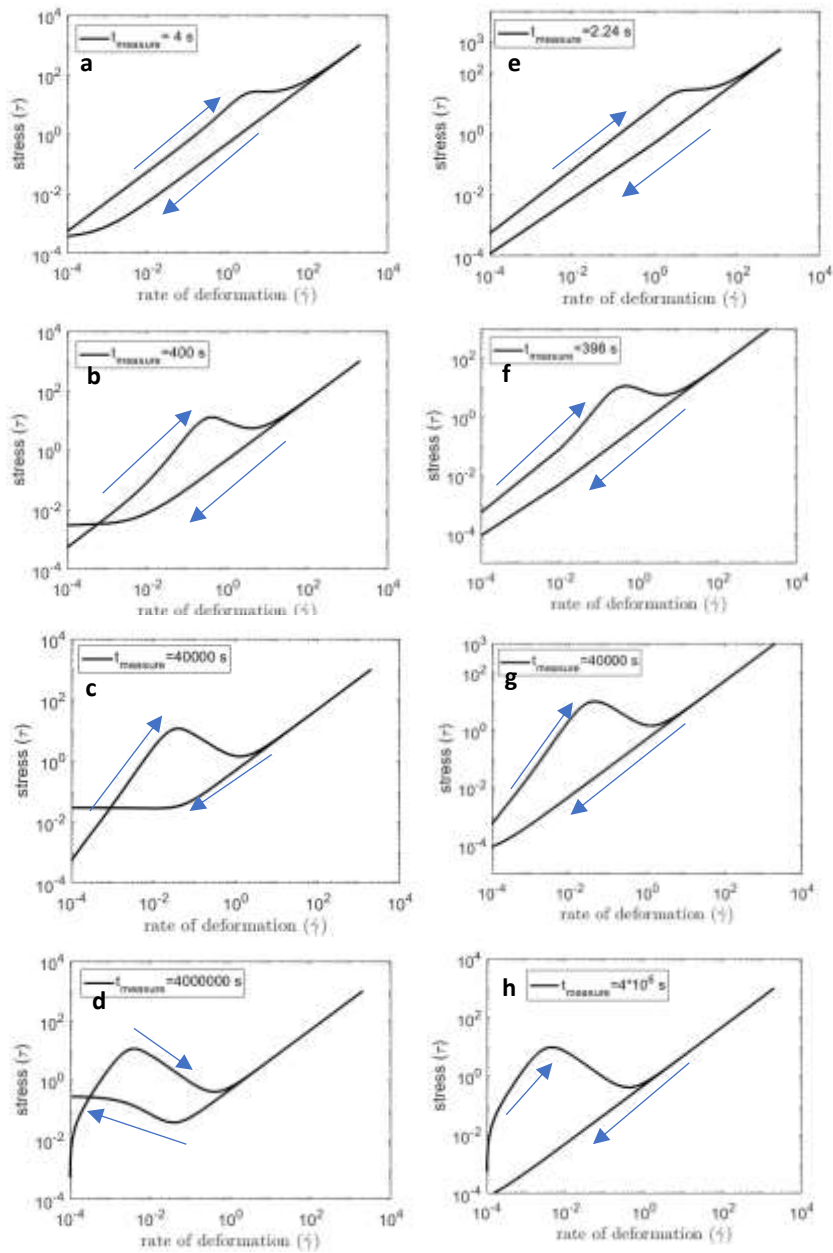

Fig. 9. Stress as a function of the rate of deformation for different values of calculation (experiment) time, for thixotropic materials with thixotropic time scale $T_0 = 100$ s, (a) $t_{measure}$=4 s (b) $t_{measure}$=400 s, (c) $t_{measure}$=40000 s, and (d) $t_{measure}$=4000000 s, and for irreversible thixotropic materials, $T_0 \rightarrow$ infinity (i.e., $\lambda_e = 0$ ), (e) $t_{measure}$=2.24 s, (f) $t_{measure}$=398 s, (g) $t_{measure}$=40000 s, and (h) $t_{measure}$=4000000 s, while parameters m=100, $\mu_s$=0.5 Pa s, $\mu_g$=5 Pa s, $G_0$=5000 Pa are identical for all cases
.

Figure 9 shows the stress as a function of shear rates in the cyclic shear-rate sweep test for irreversible and reversible thixotropic materials. Figs. 9a -9d show the effect of measurement time on the hysteresis characteristic for reversible thixotropic material with the thixotropic time scale $T_0$=1000 s. As the measurement time increases, both up and down sweeps curve characteristic changes. During up-sweep, a slower change in the shear rate provides an opportunity for more elastic deformation before the shear rate becomes higher, leading to higher stress for a lower shear rate (Fig. 9d). Similarly, the slower changes in the shear rate during shear down measurement provides enough time for material aging, leading to microstructure build-up. This leads to an increase in the stress at lower shear rates, where elastic deformation starts dominating during aging. Figure 9a is plotted where the shear rate has been increased (total time for up and down measurement is 4 s) much lower than the thixotropic time scale (1000 s). Hence, neither material gets time for elastic deformation during up-sweep nor for aging during down-sweep. Thus, a similar stress profile with a smaller hysteresis area is obtained during both up and down sweeps. As the measurement time increase, the extent of hysteresis increases (Figs. 9b and 9c). Furthermore, Figs. 9e-9h show the effect of measurement time on the characteristic of hysteresis of irreversible thixotropic materials. In this case, we observed that as measurement time increases, it only affects up-sweep, not down-sweep. During an up-sweep longer time duration at a lower shear rate cause more elastic deformation before a higher shear rate break the structure, leading to higher stress for the curve which takes the longest time of measurement (Fig. 9h). However, during the down-sweep calculation the effect of measurement time is negligible as irreversible thixotropic material is non-aging. Hence, even a longer measurement time does not affect the microstructure of the material during the down-sweep. Therefore, for all cases, similar down-sweep stress behaviours are observed. For Fig. 10, the measurement time is fixed, however, the maximum shear rate differs in each case. A lower maximum shear rate means the material has undergone more initial elastic deformation. Our shear hysteresis is qualitatively consistent with the type of shear hysteresis discussed by Radhakrishnan et al.(89)

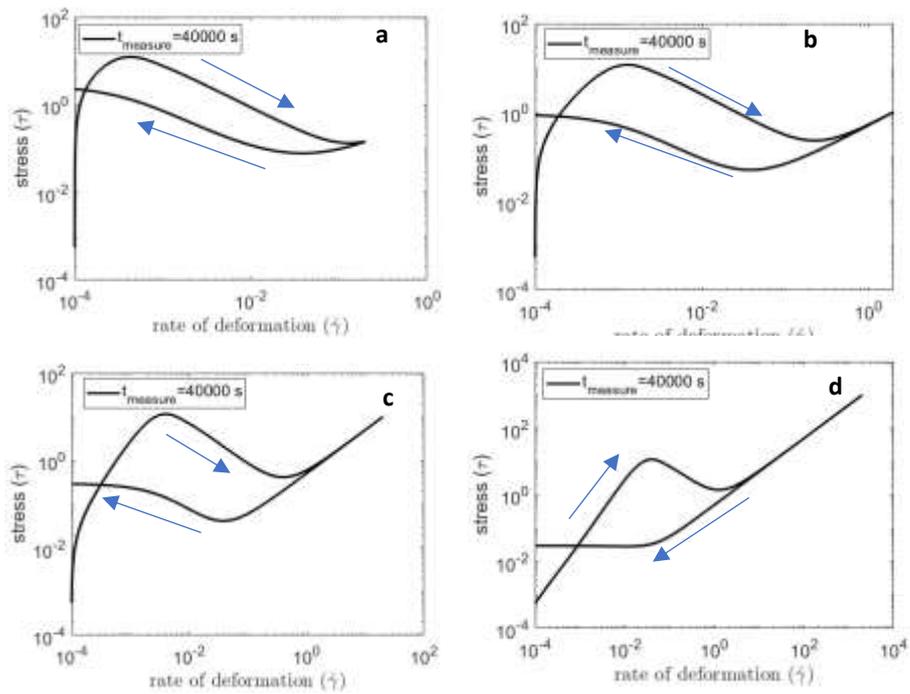

Fig. 10. Stress as a function of the deformation rate for the same calculation (experiment) time while the maximum shear rate is different using parameters $T_0 = 100$ s, $m=100$, $\mu_s=0.5$ Pa s, $\mu_g=5$ Pa s, $G_0=5000$ Pa.

Fig. 11 compares the experimental results of Mendes et al.(60) with that of our modelling prediction. From earlier results, it is clear that the duration and nature of the shear ramp also influence the hysteresis curve significantly. Hence, we selected the same shear ramp duration as used in experiments. Fig. 11a is for waxy crude oil cooled to a lower temperature (until 4°C at a cooling rate of -1°C/m). Fig. 11b is for waxy crude oil cooled to a higher temperature (until 23°C at the same cooling rate of -1°C/m). A lower final temperature produces a stronger gel, which also appears irreversible. Irreversibility is clear from the shear rate down experiments which show monotonically decreasing stress and no aging. However, weaker gel seems to be aging material at low shear rates. The softer gel shows shear banding during both shear up and down tests, whereas the stronger gel only shows shear banding during the shear up test. This indicates that while both material break during the shear-up test, only weaker waxy crude oil gel shows aging behaviour. We have used these pieces of information to predict the above results from our model. For Fig. 11c infinitely large value of thixotropic time is chosen so that the gel will not recover. However, for Fig. 11d, a large but finite value of thixotropic time, $T_0=12000$ s, is selected. We are able to qualitatively predict the experimental results. Our model initially shows an elastic jump, which does not appear in the experiment results. This depends on the measurement protocol used during experiments, in the case initial results are not recorded, then the experimental observation of initial elastic behaviour becomes difficult to notice. Apart from the initial elastic jump, both experimental and modelling results show that stress increases to a maximum before considerable breakage reduces the stress. Finally, at a large value of shear rate, both model and experiments show an increase in stress dominated by the viscous effect. During the down sweep test, the material without aging (irreversible fluids) shows a monotonic decrease in the stress, however, aging material stress increases at the lower strain rate. This happens as aging strengthens gel, and elastic stress grows at lower shear rates.

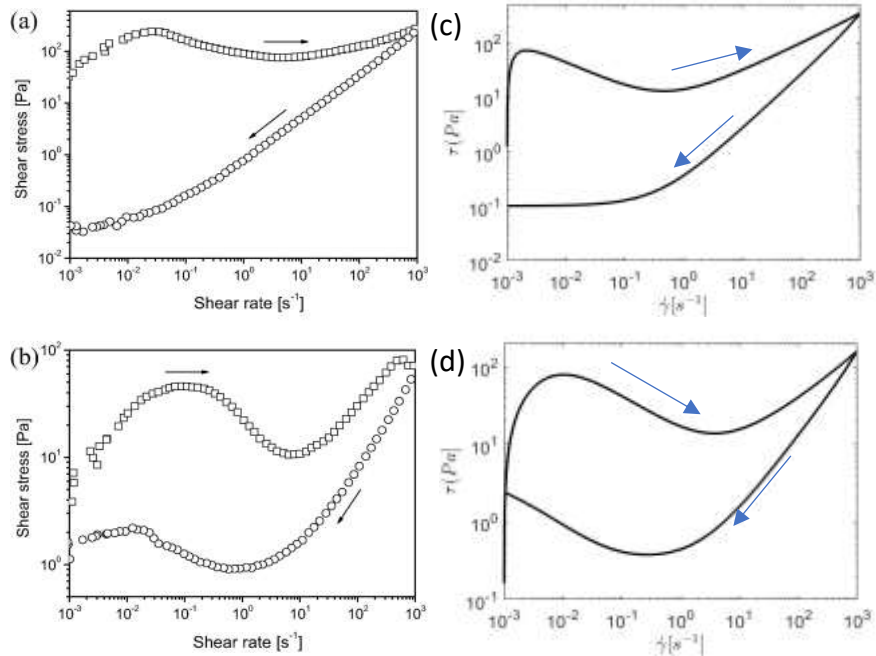

Fig. 11. Experimental results of Mendes et al.(60) showing stress as a function of the strain rate in shear sweep test (a) irreversible breakage with no aging (waxy crude oil after static cooling from 60°C to 4°C at -1°C/m and holding time 20 min), (b) reversible breakage with aging (waxy crude oil after static cooling from 50°C to 23°C at -1°C/m and holding time 20 min), (c) our model prediction corresponding to experimental results in (a) using $T_0 \to \infty$, $\mu_s$=0.05 Pa s, $\mu_g$=400 Pa s, m=40, $G_0$=15000 Pa, and (d) our model prediction corresponding to experimental results in (b) using $T_0$=12000 s (due to reversibility), $\mu_s$=0.05 Pa s, $\mu_g$=50 Pa s, m=5, $G_0$=2000 Pa

Next, we compared our shear hysteresis results with the results of Serial et al.(59). They used the milk microgel suspension for their study. For the results shown in Fig. 12a, they first increased the shear rate from low to high and then again to a low value. They performed a reverse test for another sample, i.e., high-to-low-to-high again. Their results show that the milk microgel suspension is a non-ageing-yielding material, which shows an initial elastic jump in the stress. As discussed earlier, our model predicts elastic jump at initial times (Fig. 12d). However, here we have not plotted the first few data to show a closer match with the experimental results. This is consistent with the experimental measurement, where either measuring initial data is difficult or ignored because it is error-prone due to system inertia etc. For these non-aging yielding fluids, a constant value dynamic structure parameter $\lambda_e = 0.0001$ is used (this is equivalent to taking $T_0 \sim 10^7$ s at the lower shear rate used in this experiment). A very high thixotropic time scale will also show non-aging as the micro-structure cannot be reformed during experiments. Other parameters m, $G_0$ are generally obtained from the shear rate start-up test, here it is taken to match maximum stress. Using the same parameters, we have calculated stresses in the reverse cycle (Fig. 12d), corresponding to experimental results shown in Fig. 12b. Our modelling and Serial et al. experimental results match very well.

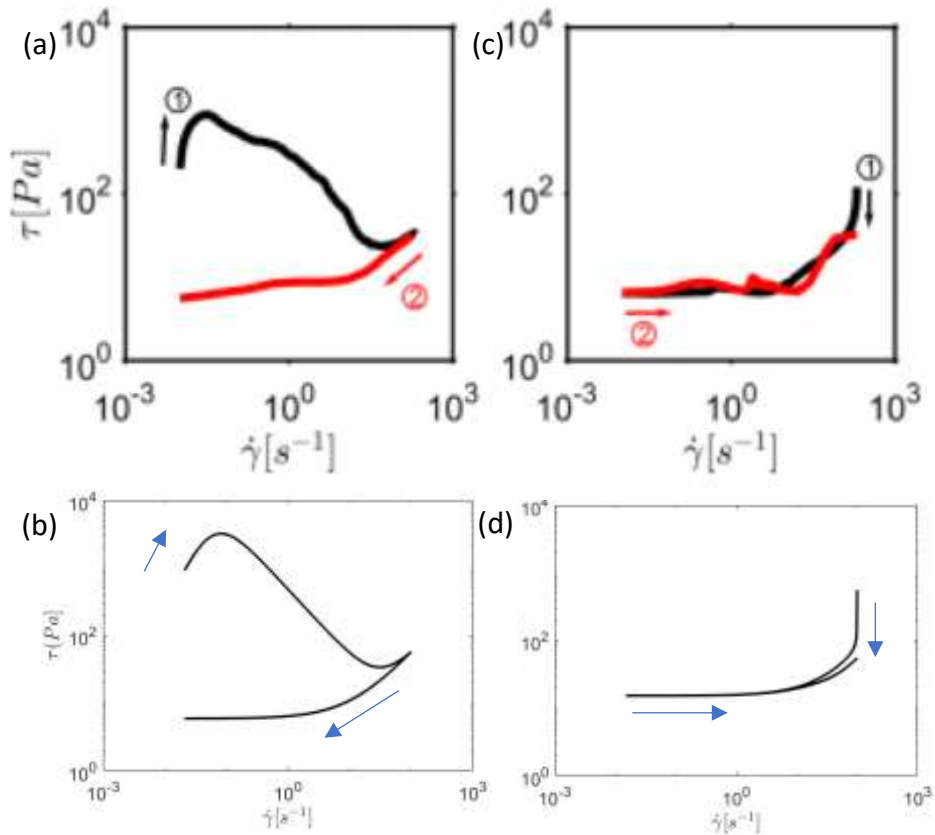

Fig. 12. Experimental results of Serial et al.(59) in shear sweep test for the milk microgel suspension (a) for shear rate low-to-high-low value, (b) for high-low-high value (c) our model prediction corresponding to experimental results for shear rate low-to-high-low value, and (d) for shear rate high-low-high value using $T_0$=100000000 s, $\mu_s$=0.5 Pa s, $\mu_g$=5 Pa s, m=0.3, $G_0$=5000 Pa.

## 4.6 Apparent viscosity as a function of the shear rate

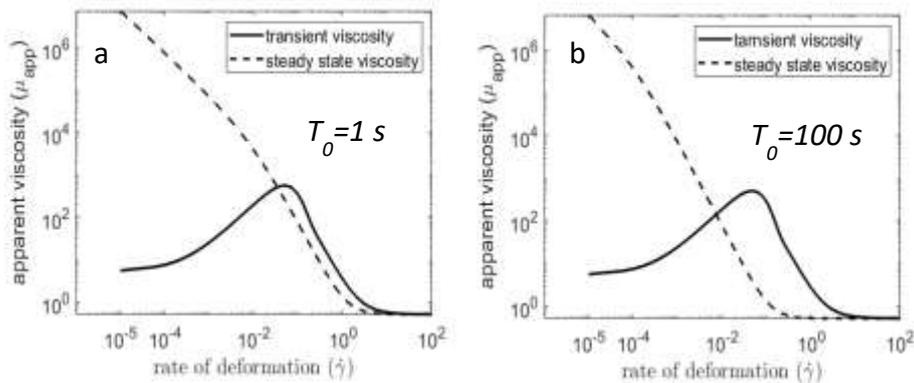

Fig. 13. Apparent viscosity as a function of the rate of deformation for different values of the thixotropic time scale (a) T$_0$= 1 s, and (b) T$_0$= 100 s other parameters $\mu_s$=0.5 Pa s, $\mu_g$=100 Pa s, m=100, G$_0$=500 Pa, are kept constant for both cases.

Figure 13 shows the apparent viscosity as a function of the shear rate for different values of the thixotropic time scale. The solid line shows the predicted viscosity when the shear rate continuously increases linearly from a low value of $10^{-5}$ s$^{-1}$ to $10^{3}$ s$^{-1}$ in 1000 s. Once the strain rate reaches the maximum value, the shear-rate down calculation is performed such that the material reaches steady state viscosity for each shear rate. The results corresponding to the shear rate-up test explain the observation of Barnes & co-workers(18,20). Barnes & co-worker, in their popular articles, claimed that everything flows and yield stress is a myth. To conclude, they have shown finite viscosity at low shear rates. Hence, they concluded that yield stress in a measurement artefact and everything flow under stress, maybe with an unobservable shear rate. Recently, Mollar et al.(19) have done a number of experiments and shown that in the case of applied stress lower than yield stress, a longer time results show an increase in the apparent viscosity with time, and eventually, it becomes very large. The increasing apparent viscosity with the waiting for is referred to as an indication of flow stop and the existence of yield stress. Hence, they concluded that yield stress is a reality and suggested fellow researchers perform the experiments for a longer time to conclude the existence of yield for a given material. Our model predicts a viscosity plateau for small shear rates when the shear rate increases continuously, similar to Barnes & Co-worker observations. At the same time, our model predicts diverging zero shear rate viscosity at the steady state condition, similar to Moller et al. observations. Hence, we can conclude that our model is capable of predicting both transient and steady-state behaviours accurately. For waxy crude oil, our modelling of a sudden drop in shear rate test results matches well with experimental results.

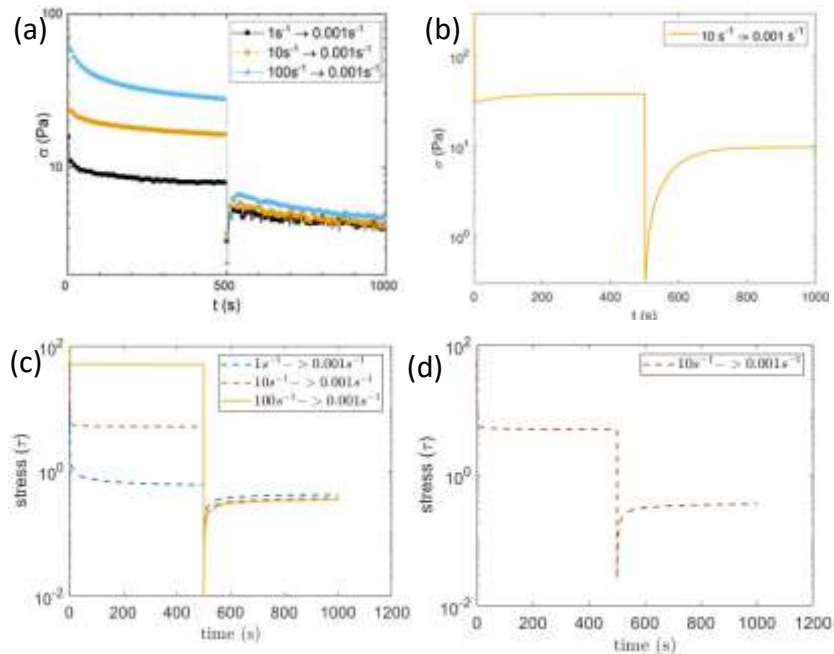

Fig. 14. Stress as a function of time in sudden step-down shear rate tests (a) experimental results Datta et al.(61), (b) modelling results of Wang and Larson(7), (c) our model results using $\mu_s$=0.5 Pa s, $\mu_g$=5 Pa s, m=100, $G_0$=5000 Pa, and $T_0$= 3000 s, in a case where initial strain rates are high for 500 s and suddenly reduced to a low value of $10^{-3}$ s.

Finally, we want to analyse the rheological response of our model against a step-down shear experiment with a sudden drop of shear rate at 500 s reported by Datta et al.(61) (Fig. 14a). They reported an initial sharp but smooth drop in stress at constant high shear rate. However, after a certain time, they suddenly dropped the shear rate from a high value to a low value and observed the stress response. Following a step-down in the strain rate, their sample showed a sudden drop in stress. The stress recovers fast, to a certain extent, when the step-down strain rate is maintained, however, after some time, the recovery becomes gradual and never reaches the original stress level. Finally, their experimental result shows a decrease in stress for a longer time when a low level of shear rate is maintained. Furthermore, Wang and Larson(7) predicted this rheological behaviour using their model (Fig. 14b), consisting of an elasto-plastic stress and a smoothly decreasing modulus near a solid boundary. Their model predicts a sharp initial decrease in stress at a high shear rate, nearly at a single shear rate, compared to the sharp but continuous decrease reported by Datta et al. and predicted by our model (Figs. 14c and 14d).

Furthermore, their model appears to predict a small increase in stress while the strain rate remains high, compared to experimental results and our prediction, where stress continues to decrease but at a much slower rate. The viscous effect at the initial stage makes the decrease in stress smoother for our model. We have included the viscous effect during the initial stage as well, this is consistent with the Oldroyd observation that in the case of quasi-static only, the elastic deformation can explain pre-yielding flow behaviour. Here, the system initially subjected to a high or moderate shear rate is far from a quasi-static condition, and hence the inclusion of viscous dissipation helps our model in predicting better initial behaviours. Furthermore, Wang and Larson's model predicts slower recovery during step-down shear rate tests compared to experimental observation and our model results. Both our model and Wang and Larson's model were unable to predict a long-time decrease in stress. Wang and Larson attributed this to the lack of more complex long-time dynamics of actual material. They recommended the inclusion of multiple models of structure parameters λ to capture the long-time behaviour of material after step-down experiments, as suggested by Mewis and Wanger(55). However, we observed that at a low shear rate, aging becomes more than the building term for broken gel, and hence we are unable to predict a further reduction in the stress. Using our model, the same can be predicted when a final lower shear rate is also in the range of shear rejuvenation dominant regime, not aging dominant regime.

## 4.7 Non-monotonic viscosity bifurcation and delayed yielding during creeping flow

Here, we performed a quasi-static analysis of strain and apparent viscosity evolution as a function of time for a constant applied stress condition. The detailed transient start-up flow needs a separate treatment. Under quasi-static conditions, when the applied stress is less than the yield stress, our model in Eq. 18 predicts no steady-state flow. 1D momentum balance equation between two parallel plates can be written as,

$$\rho \frac{\partial u}{\partial t} = \frac{\partial \tau}{\partial y} \qquad (35)$$

In quasi-static conditions above equation can be approximated as

$$\frac{\partial \tau}{\partial y} = 0 \qquad (36)$$

Where stress is given by Eq 18b. For $\lambda_e = 0$ in the case of a constant applied stress, the solution of the above equation reduces to

$$\tau_{applied\ stress} = \mu_s \dot{\gamma} + \mu_g \left(\frac{1}{(2m\gamma+1)^{1/2}}\right)\dot{\gamma} + G_0 \left(\frac{1}{(2m\gamma+1)^{3/2}}\right)\gamma \quad (37)$$

Fig. 15 shows the strain and the apparent viscosity as a function of time for applied stress less than, equal to, and more than the static yield stress $\left(\tau_{ys} = \frac{G_0 \gamma_{yield}}{3^{3/2}}\right)$. The applied stress lower than the static yield stress in Eq. (37) results in a finite small strain at steady state, as shown in Fig. 15a (i.e., for applied stress $1/3\tau_{ys}$, $2/3\tau_{ys}$, and $\tau_{ys}$). Hence, it can be concluded that no steady-state flow is observed for the applied stress less than or equal to the yield stress. In these cases, the applied stress is balanced by the elastic force. However, in the case of applied stress slightly higher yield stress, i.e., $1.01\tau_{ys}$, an increasing strain with time is obtained, indicating yielding and steady-state flow. A further increase in applied stress results in faster deformation, and immediate yielding can be observed. Fig. 15b shows the apparent viscosity as a function of time, which is consistent with the strain function. The apparent viscosity diverges for applied stress lower than yield stress, however, for applied stress larger than the yield stress, the apparent viscosity becomes significantly low at a steady state.

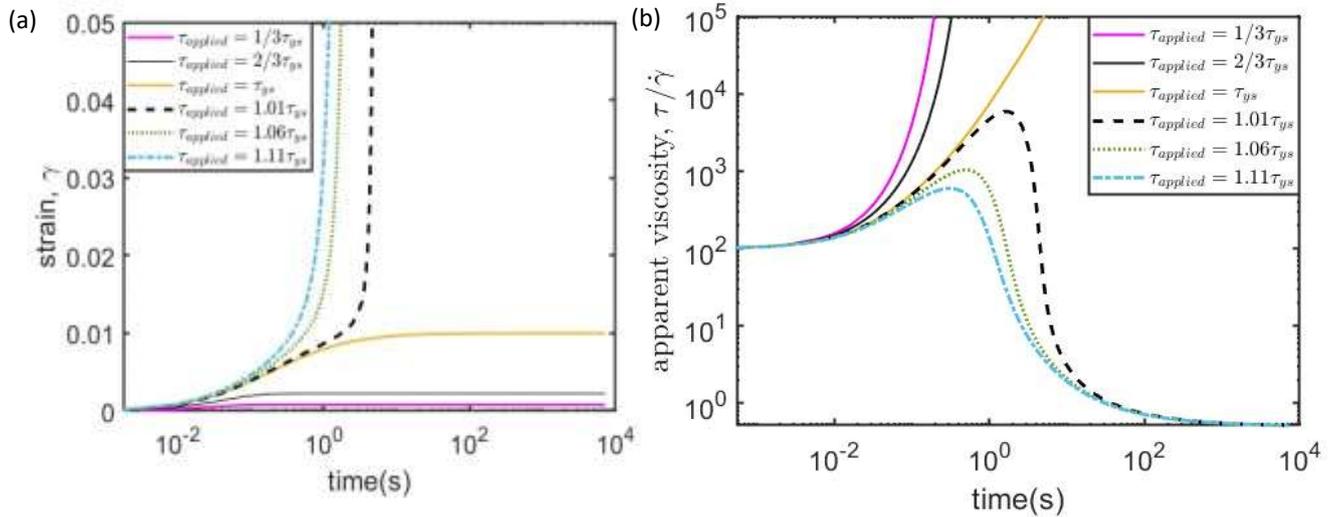

Fig. 15. (a) Strain, and (b) apparent viscosity as a function of time for different applied stresses for $\mu_s$=0.05 Pa s, $\mu_g$=100 Pa s, m=100, and $G_0$=5000 Pa.

For some structured fluids, it has been observed that when the applied stress is less than the initial static yield stress, the deformation remains less than the corresponding yield strain for a long time. However, for some materials, it is observed that suddenly deformation increases beyond yield strain, and flow starts. This is referred to as delayed yielding. In Eq. (5), we have seen earlier that when the applied stress is not enough to induce deformation beyond the yield point, the second exponential term reduces the structure parameter values. The second exponential term is a function of time and is influenced by the thixotropic time scale. Hence, the yield stress requirement also reduces. However, the structure parameter reduction also depends on the build-up term. When the shear rate approaches zero, the build-up term dominates in the aging material, and the microstructure becomes stronger over time. However, for non-aging materials like waxy crude oil, we can consider a constant value of $\lambda_e$. In such cases, the build-up can be neglected, resulting in a constant first term, and the second term reduces with time. This results in lower yield stress requirement, and strain can slowly become more than yield strain. Once deformation in the material crosses the static yield strain value, the elastic stress requirement for flow decreases. This increases the deformation, leading to a further reduction in the second term of the structure parameter. This causes an increase in the strain rate, and the equilibrium structure parameter also reduces, and suddenly deformation rate starts increasing rapidly. To incorporate this effect, we have substituted Eqs. (10) and (16) in Eq. (17) to formulate a constitutive equation corresponding to first and third-order gel degradation with the possibility of a delayed restart, respectively.

For the first-order gel degradation model constitutive equation with the possibility of delayed yielding becomes

$$\tau = \mu_s \dot{\gamma} + \mu_g \left(\lambda_e + (1-\lambda_e)e^{-m\gamma}e^{-t/T_0}\right)\dot{\gamma} + G_0 \left(3 * \lambda_e e^{-2m\gamma} e^{-2t/T_0} + e^{-3m\gamma} e^{-3t/T_0}\right) \gamma$$
(38)

Similarly, for third-order gel degradation kinetic rheological model for thixotropic elasto-visco-plastic material with the possibility of delayed yielding becomes as follows.

$$\tau = \mu_g \left(\lambda_e + \frac{e^{-t/T_0}}{(2m\gamma+1)^{1/2}}\right)\dot{\gamma} + \mu_\infty \dot{\gamma} + G_0 \left(\frac{3*\lambda_e*e^{-2t/T_0}}{(2m\gamma+1)} + \frac{e^{-3t/T_0}}{(2m\gamma+1)^{3/2}}\right)\gamma \qquad (39)$$

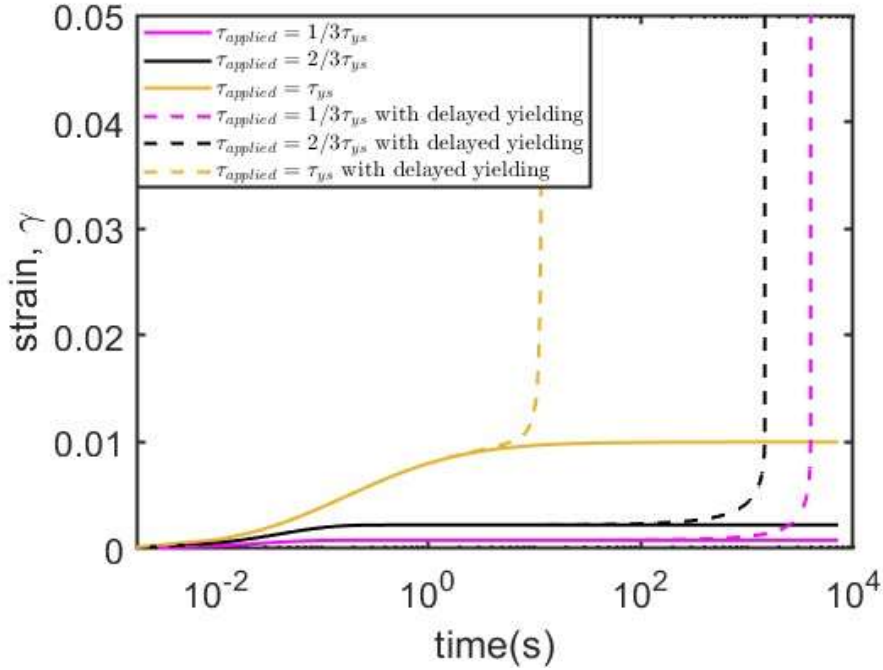

Fig. 16. Strain as a function of time using with and without a delayed yielding model for different applied stresses for $\mu_s$=0.05 Pa s, $\mu_g$=100 Pa s, m=100, $G_0$=5000 Pa, and $T_0$= 3600 s.

We have substituted stress expression from Eq. 39 to Eq. 37 and solved for a constant value of applied stress less than yield stress. Fig. 16 shows that a constant strain value is obtained at a large time without a time-dependent term in the constitutive equation. This indicates a balance between the applied stress and elastic force, resulting in no steady-state flow. However, the inclusion of a time-dependent term in the constitutive equation, as in Eq. 39, decreases the stress requirement. Fig. 16 shows that a delayed flow start is observed in the case of Eq. 39. In this case, $T_0$=1 hr is used. For $T_0$=1 hr, when the applied stress is the same as yield stress flow quickly starts (~10 sec). However, a long delay (~hr) in flow is observed for a lower value of applied stress. Hence, it can be concluded that for delayed flow start, the magnitude of applied stress and duration of the applied stress both are important. When the applied stress is close to the yield stress, flow begins sooner than when the applied stress is much lower than the yield stress.

## 5  Conclusion:

In this work, we have discussed the effect of structure degradation kinetics on the prediction capability of the corresponding rheological model. Most previous works argue that either shear rate or shear stress is responsible for gel breakage. However, in the case of applied stress conditions also gel responds with the development of shear rate, including in the true yielding materials with applied stress lower than the yield stress. Hence, the model consisting of either shear rate-dependent breakage or shear stress breakage has to overcome yield strain. Therefore, we converted our gel structure parameter into strain dependent model, which has information on stress and shear rate histories. The conversion from shear rate-dependent gel degradation to strain-dependent structure parameter becomes evident in the case of the Kee et al.(63) model. Hence, we argue that in Kee et al.'s model, we can select the dynamic structure parameters to get all the models mentioned in Table 1. Further, it can be noticed that shear rate dependence in the rheological parameters comes from dynamic equilibrium structure parameters. Thus indicate that breakage can depend on the total deformation alone, however, the net change in the structure on shearing strength.

 We have formulated a new rheological model based on structure parameters. Our rheological model is a simple algebraic equation, requiring only four parameters for the irreversible TEVP model and five for the TEVP model, compared to the existing one requiring up to thirteen parameters. Despite being an algebraic equation with fewer parameters, our model can qualitatively capture many rheological behaviours of reversible and irreversible thixotropic elasto-visco-plastic material, which were earlier pointed out to be challenging to model.

Our rheological model explains both viscosity plateau at low shear rates and diverging zero-shear rate viscosity. The above observation is one of the most controversial observations in thixotropic rheology. It also determines if the material has true yield stress or not. As pointed out in the literature, either of the above conclusions can be made depending on the patience of the experimentalist. Similarly, our model predicts a viscous plateau for continuously varying shear rates (implying no true yield stress) and diverging zero-shear rate viscosity at steady state conditions (hence true yield stress material). Depending on the material's shear histories and characteristics of the dynamic equilibrium constant value, $\lambda_e$. The choice of $\lambda_e$ reduces our model to either Bingham, Herschel Bulkley type or Newtonian model in the steady state conditions.

Our model also predicts experimentally observable transient shear banding due to microstructure breakage by shear rejuvenation and steady-state shear banding due to aging. During the transient flow of virgin gel, our model qualitatively shows similar behaviour to the steady state behaviour of the de Souza Mendes model (43) and predicts similar results as Dimitriou and McKinley(46) at steady state conditions. It is also able to explain different types of shear hysteresis during shear rate cyclic test, waiting time-dependent stress overshoots during a constant shear rate flow start-up, and sudden step down in shear rate test results effectively. Finally, we are also able to explain different flow behaviours during shear rate start-up flow, i.e., initial elastic jump, creeping flow, the subsequent increase in stress until maxima, followed by a decrease in stress, and finally, the stress increases as a function of the shear rate. It can also explain delayed yielding phoneme, which is challenging to model using the existing model. Hence, our model will be useful in explaining several rheological phenomena that are otherwise difficult to explain using same model. At the same time, our model can be improved further, especially for predicting the results of step-down stress cases, by identifying the recoverable and unrecoverable deformation.

# 6 Acknowledgment

We acknowledge financial support from the Science and Engineering Research Board, Government of India. We also thank the anonymous reviewers of "Tikariha and Kumar. *Journal of Fluid Mechanics* 911 (2021): A46 &., Tikariha and Kumar. *Journal of Non-Newtonian Fluid Mechanics* 294 (2021): 104582." For pointing out important aspects of the rheological model developed by Kumar et al." *Journal of Non-Newtonian Fluid Mechanics* 231 (2016): 11-25.", which results in the present work. Kumar et al. work is a special case of present work that can be obtained by putting $\lambda_e=0$.

Reference,